\documentclass[journal,compsoc]{IEEEtran}

\usepackage{blindtext, graphicx, amsmath, amssymb, booktabs}
\usepackage{float}
\usepackage[utf8]{inputenc}
\usepackage[english]{babel}
\usepackage{algorithm}
\usepackage{algorithmic}
\usepackage{color}
\usepackage{verbatim}
\usepackage{esvect}
\usepackage{amsmath,amsfonts,amssymb}

\usepackage{acronym}
\usepackage{textcomp}
\usepackage{stfloats}
\usepackage{url}
\usepackage{algorithm,algorithmic}
\usepackage{pgfplots}
\usepackage{tikz}
\usetikzlibrary{calc}
\makeatletter
\newcommand{\gettikzxy}[3]{%
  \tikz@scan@one@point\pgfutil@firstofone#1\relax
  \edef#2{\the\pgf@x}%
  \edef#3{\the\pgf@y}%
}
\usetikzlibrary{spy,backgrounds}
\newcommand{\comm}[1]{}

\newcommand{\ignore}[1]{}
\ifCLASSOPTIONcompsoc
  \usepackage[nocompress]{cite}
\else
  \usepackage{cite}
\fi

\newcommand{\red}[1]{{\color{red}{#1}}} 
\newcommand{\brown}[1]{{\color{brown}{#1}}} 
\newcommand{\cyan}[1]{{\color{cyan}{#1}}} \newcommand{\violet}[1]{{\color{violet}{#1}}} 

\long\def\comment#1{}

\DeclareMathOperator*{\argmin}{arg\,min}

\newfont{\bbb}{msbm10 scaled 700}

\newfont{\bb}{msbm10 scaled 1100}

\newcommand{\av}{{\bf a}}
\newcommand{\bv}{{\bf b}}
\newcommand{\cv}{{\bf c}}
\newcommand{\dv}{{\bf d}}
\newcommand{\ev}{{\bf e}}

\newcommand{\hv}{{\bf h}}

\newcommand{\nv}{{\bf n}}

\newcommand{\pv}{{\bf p}}
\newcommand{\qv}{{\bf q}}

\newcommand{\sv}{{\bf s}}
\newcommand{\tv}{{\bf t}}

\newcommand{\wv}{{\bf w}}

\newcommand{\yv}{{\bf y}}

\newcommand{\Gm}{{\bf G}}

\newcommand{\etav}{{\boldsymbol\eta}}

\newcommand{\muv}{\hbox{\boldmath$\mu$}}
\newcommand{\zetav}{\hbox{\boldmath$\zeta$}}

\newcommand{\omegav}{\hbox{\boldmath$\omega$}}
\newcommand{\xiv}{\hbox{\boldmath$\xi$}}

\newcommand{\varthetav}{\hbox{\boldmath$\vartheta$}}

\newcommand{\Sigmam}{\hbox{\boldmath$\Sigma$}}

\newcommand{\Omegam}{\hbox{\boldmath$\Omega$}}

\renewcommand{\arg}{{\hbox{arg}}}

\acrodef{bs}[BS]{base station}
\acrodef{ue}[UE]{user equipment}
\acrodef{d2d}[D2D]{device-to-device}
\acrodef{mle}[MLE]{maximum likelihood estimation}

\begin{document}
\bstctlcite{IEEEexample:BSTcontrol}
%

\title{Zero Access Points 3D Cooperative Positioning via RIS and Sidelink Communications}

\author{
Mustafa~Ammous,~\IEEEmembership{Student~Member,~IEEE},
Hui~Chen,~\IEEEmembership{Member,~IEEE}, Henk~Wymeersch,~\IEEEmembership{Senior~Member,~IEEE},
and~Shahrokh~Valaee,~\IEEEmembership{Fellow,~IEEE}

\IEEEcompsocitemizethanks{\IEEEcompsocthanksitem M. Ammous and S. Valaee are with the Department of Electrical and Computer Engineering, University of Toronto, Toronto, ON M5S 3G4, Canada. E-mail:mustafa.ammous@mail.utoronto.ca; valaee@ece.utoronto.ca.
\IEEEcompsocthanksitem H.~Chen and H.~Wymeersch are with the Department of Electrical Engineering, Chalmers University of Technology, 412 58 Gothenburg, Sweden. E-mail: \{hui.chen; henkw\}@chalmers.se).}
\thanks{The work of M. Ammous and S. Valaee was financially supported by Huawei Technologies Canada Inc.. The work of H. Chen and H.~Wymeersch was supported by the EU H2020 RISE-6G project under grant 101017011.}}


\IEEEtitleabstractindextext{%

\begin{abstract}
Reconfigurable intelligent surfaces (RISs) are expected to be a main component of future 6G networks, due to their capability to create a controllable wireless environment, and achieve extended coverage and improved localization accuracy. 
In this paper, we present a novel cooperative positioning use case of the RIS in mmWave frequencies, and show that in the presence of RIS, together with sidelink communications, localization with zero access points (APs) is possible. We show that multiple (at least three) half-duplex single-antenna user equipments (UEs) can cooperatively estimate their positions through device-to-device communications with a single RIS as an anchor without the need for any APs. We start by formulating a three-dimensional positioning problem with Cramér-Rao lower bound (CRLB) derived for performance analysis. 
After that, we discuss the RIS profile design and the power allocation strategy between the UEs.
Then, we propose low-complexity estimators for estimating the channel parameters and UEs' positions. Finally, we evaluate the performance of the proposed estimators and RIS profiles in the considered scenario via extensive simulations and show that sub-meter level positioning accuracy can be achieved under multi-path propagation.  
\end{abstract}

}

\maketitle

\IEEEdisplaynontitleabstractindextext
%
\IEEEpeerreviewmaketitle

\section{Introduction}
\subsection{Motivation}
The requirements of communication networks, such as high data rate, connectivity, and reliability, are expected to increase with the growing interest in services and technologies such as Internet-of-Things (IoT), intelligent transportation systems (ITS), and smart cities \cite{6G-use-cases}. To meet these requirements, 5G and beyond 5G (B5G) networks are expected to utilize higher bandwidths, higher frequencies, massive multiple-input multiple-output (mMIMO)\cite{general_comm, irs1}, which are key factors for enabling joint communication and sensing on a single platform in future networks \cite{jcas-survey}.
One of the challenging issues for the systems operating at high frequency (e.g., mmWave and THz bands) is the high penetration and path losses, limiting the coverage of communication and sensing services. 
Network densification, by adding more \acp{bs} or access points (APs), helps to improve the connectivity and coverage in the network but that will increase the hardware and deployment costs, and energy consumption drastically \cite{irs2}. Instead, adding passive nodes to the network will achieve a more optimized system cost and lower energy consumption. 

Recently, the concept of {\it reconfigurable intelligent surfaces} (RISs), also referred to as {\it intelligent reflective surfaces} (IRSs), has drawn attention from both academia and industry, with the advantage of relatively low-cost and the capability to improve coverage and reduce energy consumption \cite{irs3}. RIS is a digitally-controlled planar array consisting of a large number of low-cost metamaterial elements that reflect incident signals \cite{irs4}. RISs can improve communications by creating an indirect path between a BS and a user equipment (UE) when the line-of-sight (LoS) is not available, or to boost the communication capability in the presence of LoS \cite{control}. 
In addition to improving communications, it has been shown that RIS can assist localization\footnote{We use the terms localization and positioning interchangeably in this paper, as a core function of sensing, indicating the estimation of UE location (or position).} services \cite{cm-ris-location}. Although one BS is sufficient for communication purposes, usually more anchors (BSs) are needed for location estimation~\cite{RIS_magazine}. For example, three time-difference-of-arrival (TDOA) measurements from four BSs are needed for localization in 4G systems while angle measurements from two BSs (equipped with antenna arrays) are sufficient for localization in 5G networks \cite{control}. By acting as an additional anchor, RIS enables location estimation with only one BS \cite{control}. Recently, it has been shown that with the assistance of RIS, a full-duplex UE can achieve self-localization without the need for any APs or BSs \cite{zero-ap-ris}.  

Furthermore, the introduction of sidelink or \ac{d2d}\footnote{Sidelink or D2D communications refer to direct communications between UEs without the data passing through the AP.} communications in 5G-NR opens the road for a slew of opportunities to achieve more accurate UE positioning due to the availability of a larger number of radio measurements via cooperation between different UEs \cite{sidelink-intro, cp-ris-mustafa}. Without the need for a coordinating AP, the UEs can autonomously select sidelink resources from a pre-configured sidelink resource pool(s) \cite{d2d-standard}. The introduced sidelink communication can thus make localization service available in partial-coverage and out-of-coverage scenarios such as indoor UEs or vehicles in tunnels \cite{RIS_magazine}. In this paper, we argue that with D2D communications and the cooperation between multiple UEs, a single RIS anchor can enable AP-free positioning. 


\subsection{Related Work}
Several recent works have studied the importance of RISs in improving positioning accuracy \cite{irs5,zanaty-bounds,irs8,ris-fingerprint,dardari-ris}. In \cite{irs5, zanaty-bounds}, the authors derive the position and orientation error bounds in a RIS-assisted system with one BS and one UE using the Fisher information matrix (FIM). They also show that the positioning error decreases by finding the optimal configuration of the RIS elements. In \cite{irs8}, the authors optimize the configuration of RIS elements by minimizing a localization loss function with received signal strength (RSS) measurements in a multi-user scenario. It has also been shown that the RIS can be used to build a radio map of RSS measurements for localization using fingerprinting and supervised learning \cite{ris-fingerprint}. In \cite{dardari-ris}, the authors consider a positioning scenario with multiple RISs and non-line-of-sight (NLoS) links. They propose the design of orthogonal phase shifts at the RISs to be able to separate the signals arriving at the UE from different RISs for location estimation. Other studies utilize RISs for localizing and tracking objects, and creating the image of the scene \cite{tracking, SW, RIS_OTFS}. However, none of these works considers AP-free localization or tracking. 


Other works have considered cooperative positioning (CP) with RIS-assisted systems. In \cite{pimrc,cop-ris-d2d}, the authors have considered positioning systems with one BS, one RIS and multiple UEs with \ac{d2d} capability. It has been shown that measurements collected from D2D communications in addition to downlink measurements can improve positioning performance \cite{pimrc,cop-ris-d2d}. However, both works only consider direct D2D links and ignore the D2D links in the UE-RIS-UE paths. On the other hand, the authors in \cite{cop-ris-2bs} derive the performance limits of a CP system with multiple BSs, one RIS and one UE. Again, their results show improvement in positioning accuracy in comparison to non-cooperative methods, and all the mentioned works assume the availability of BS(s).

More recently, it has been argued that RISs can enable localization and mapping without the need for BSs \cite{RIS_magazine}. In \cite{zero-ap-ris}, the authors propose a new use case for the RIS where a full-duplex single-antenna UE can localize itself by leveraging controllable reflections with zero APs (i.e., acting like a radar). Again, in \cite{slam_ris_zero_ap1, slam_ris_zero_ap2}, the authors show that a full-duplex multiple-antenna UE can perform simultaneous localization and mapping of the environment with the aid of RIS and zero APs. In \cite{cp-ris-mustafa}, the authors propose a CP use case in a scenario with one RIS and two single-antenna UEs. They assume that one UE acts like a radar and transmits pilot signals to the other UE via a direct path and an indirect path through the RIS. The radar-like UE first localizes itself and then localizes the other UE. However, all the previous works that study RIS-enabled zero AP localization assume that at least one UE has a full-duplex capability which might not be available in practice.

\subsection{Contribution}
In this paper, we study a novel CP use case of the RIS with D2D communications and without the need for APs or BSs. We consider a positioning scenario with one RIS and $K$ single-antenna UEs. We build the system model based on a uniform planar array (UPA) architecture at the RIS and orthogonal frequency-division multiplexing (OFDM) signals. The system model proposed in this paper is an extension of the work in \cite{cp-ris-mustafa}, where the authors only consider two-dimensional (2D) CP with two UEs and one UE has a full-duplex capability. The main contributions of this work are:
\begin{itemize}
    \item \textbf{RIS-enabled CP problem formulation:}
    In contrast to \cite{cp-ris-mustafa}, we expand the system model to consider a three-dimensional (3D) scenario involving $K$ single-antenna UEs with unknown locations and one RIS acting as an anchor. In addition, we assume that none of the UEs has a full-duplex capability. Hence, we leverage D2D communications between the UEs to localize them. Our feasibility analysis reveals that at least $K=3$ UEs are needed to make the location identifiable. In addition, we study the performance limits of the CP system by deriving the  the Cramér-Rao lower bound (CRLB) on the channel parameters and the UEs' positions. The derived CRLB will be used to evaluate the proposed estimators as well as the proposed RIS profile designs.
    \item \textbf{RIS profile design:}
    To assist the estimation of the channel parameters, we decompose the received signal into two components, a LoS path and a RIS path, via orthogonal RIS profiles. To design the phase shifts at the RIS, we propose the use of random phase codebook and directional codebook. The choice of the codebook depends on whether prior information about the UEs' positions is available or not. Our proposed directional codebook depends on spatial frequencies rather than angles due to the unknown positions of the UEs. In addition, a generalization of the directional codebook involving a power allocation between the UEs is proposed and evaluated.
    \item \textbf{Positioning algorithm development:} 
    To perform localization of the UEs, we use a two-stage approach. First, we start by estimating the channel parameters for each pair of UEs, including delay estimations (LoS and reflected path through RIS), and spatial frequency estimations. Then, we formulate a simple one-dimensional (1D) coarse search localization algorithm, followed by \ac{mle}, to obtain the positions of all the UEs.  
    \item \textbf{Performance evaluation:} 
    We evaluate the performance of the proposed estimators in terms of positioning error via extensive simulations. We verify that the proposed estimators attain the CRLB at a sufficient transmitting power. We show that a sufficiently large RIS can replace an AP without any performance degradation, and the proposed method can achieve sub-meter level positioning even under multi-path propagation. 
\end{itemize}

The rest of the paper is organized as follows. Section~\ref{sec:problem-formulation} discusses the system setup and the channel model. Then, we derive the lower bounds of the channel parameters and positions estimation in Section~\ref{sec:crlb}. Next, we propose different RIS profile designs in Section~\ref{Sec:profile design}. We then discuss the channel parameter estimation and proposed localization algorithm in Section~\ref{sec:proposed_estimators}. Finally, we evaluate the performance of the proposed method via numerical studies in Section~\ref{sec:simulation}, followed by Section~\ref{sec:conclusion} that concludes this work.

\section{Problem Formulation}
\label{sec:problem-formulation}

\begin{figure}
    \centering
    \begin{tikzpicture}
    \node (image) [anchor=south west]{\includegraphics[width=.8\linewidth]{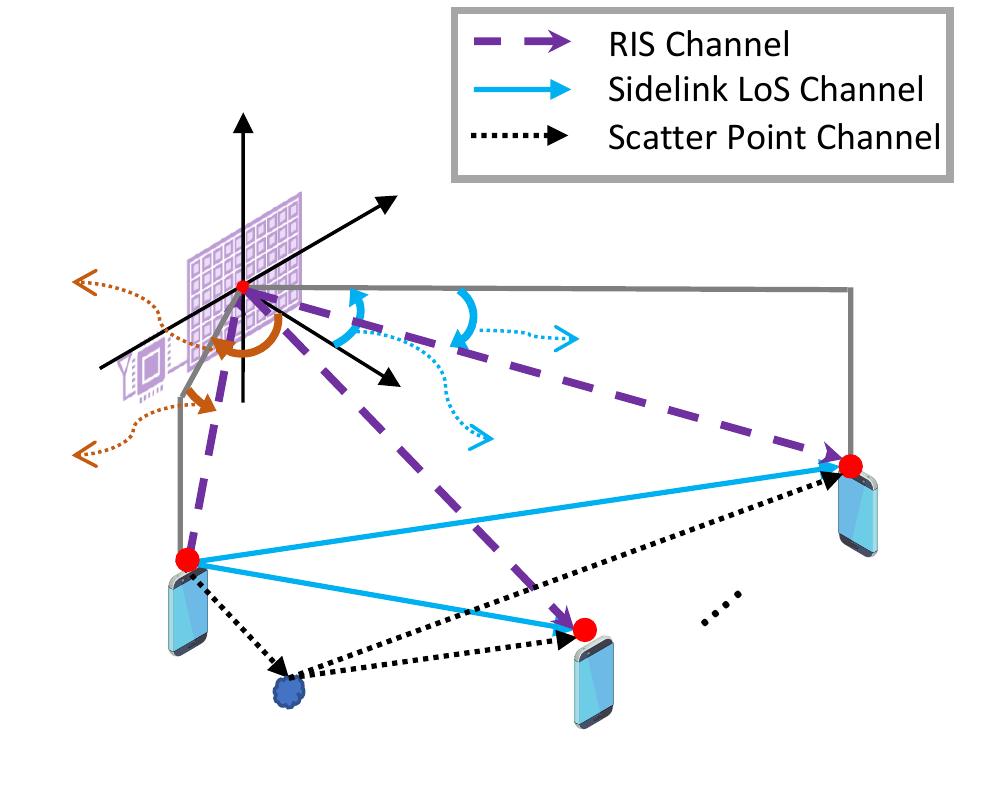}};
    \gettikzxy{(image.north east)}{\ix}{\iy};
    \node at (0.43*\ix,0.54*\iy)[rotate=0,anchor=north]{{$x$}};
    \node at (0.425*\ix,0.785*\iy)[rotate=0,anchor=north]{{$y$}};
    \node at (0.25*\ix,0.90*\iy)[rotate=0,anchor=north]{{$z$}};
    \node at (0.22*\ix,0.75*\iy)[rotate=0,anchor=north]{{\red{$\pv_\text{R}$}}};
    \node at (0.15*\ix,0.36*\iy)[rotate=0,anchor=north]{{\red{$\pv_{\text{U}_1}$}}};
    \node at (0.65*\ix,0.29*\iy)[rotate=0,anchor=north]{{\red{$\pv_{\text{U}_2}$}}};
    \node at (0.92*\ix,0.49*\iy)[rotate=0,anchor=north]{{\red{$\pv_{\text{U}_K}$}}};
    \node at (0.21*\ix,0.2*\iy)[rotate=0,anchor=north]{UE $1$};
    \node at (0.61*\ix,0.12*\iy)[rotate=0,anchor=north]{UE $2$};
    \node at (0.88*\ix,0.32*\iy)[rotate=0,anchor=north]{UE $K$};
    \node at (0.3*\ix,0.13*\iy)[rotate=0,anchor=north]{SP};
    \node at (0.04*\ix,0.7*\iy)[rotate=0,anchor=north]{{\brown{$\theta_{\text{az},1}$}}};
    \node at (0.04*\ix,0.49*\iy)[rotate=0,anchor=north]{{\brown{$\theta_{\text{el},1}$}}};
    \node at (0.56*\ix,0.5*\iy)[rotate=0,anchor=north]{{\cyan{$\theta_{\text{az},K}$}}};
    \node at (0.64*\ix,0.62*\iy)[rotate=0,anchor=north]{{\cyan{$\theta_{\text{el},K}$}}};
    \end{tikzpicture}
    \caption{The positioning system with one RIS, and $K$ single-antenna UEs showing UE $1$ as the transmitter.}
    \label{scenario}\vspace{-5mm}
\end{figure}
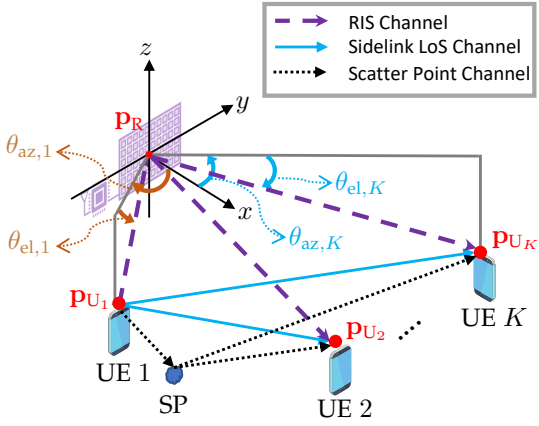
\subsection{System Setup} 
Consider the 3D scenario shown in Figure~\ref{scenario}. The  system consists of one RIS and $K\ge3$ single-antenna UEs. The RIS consists of $M$ passive reflecting elements placed in a UPA facing the $x$-axis. We assume that the location of the RIS is known where $\pv_\text{R}=[x_\text{R},y_\text{R},z_\text{R}]^\top$ represents the position of the RIS center as well as the origin of the Cartesian coordinate system. We denote the unknown position of the $k$-th UE where $k \in \{1,2, \ldots, K\}$ as $\pv_{\text{U}_k}=[x_{\text{U}_k},y_{\text{U}_k},z_{\text{U}_k}]^\top$. In addition, we assume that there are $S_j$ scatter points (SPs) for each receiving UE $j$ and the location of SP $s$ seen by the $j$-th receiving UE is denoted by $\pv_{\text{SP}_{j,s}} = [x_{\text{SP}_{j,s}},y_{\text{SP}_{j,s}},z_{\text{SP}_{j,s}}]^\top$ where $s \in \{1,2,\ldots,S_j\}$. 

We assume that the UEs communicate with each other directly by utilizing D2D communications at the mmWave band (i.e., frequencies above 24 GHz \cite{3gpp2}). As seen in Figure~\ref{scenario}, we assume that only one UE is transmitting at a time to reduce interference. We assume that the $K$ UEs are located within a small geographical area and thus we have a fully-connected D2D network similar to \cite{cop-ris-d2d, geo_area}. Each receiving UE receives the transmitted signal from three paths; a direct path, a reflected path through the RIS, and reflections from SPs in the environment.

\subsection{Channel Model}
As mentioned earlier, we assume only one UE is transmitting at a time and the other UEs are in the receiving mode. Without loss of generality, let us assume that UE $i$ is transmitting OFDM pilot signals with $N$ subcarriers for $T$ consecutive time slots. We also assume that the channel coefficients are fixed during the $T$ transmissions. Furthermore, it is assumed that all pilot symbols  transmitted by UE $i$ are equal to $\sqrt{E_i}$, where $E_i$ is the energy of the symbol. Similar to \cite{ris-loc-siso, JrCUP}, we adopt the geometric channel model to represent the received signal. Then, the received signal at UE $j$ from the transmitting UE $i$ at the discrete times $t = 1, 2, \ldots, T$ can be defined as: 
\begin{equation}
   \yv_{ij,t} = \yv_{ij,t}^{(\text{LoS})}   
   + \yv_{ij,t}^{(\text{RIS})} + \yv_{ij,t}^{(\text{MP})} +  \nv_{j,t}, \label{recived_signal}
\end{equation}
where $\yv_{ij,t}^{(\text{LoS})}$, $\yv_{ij,t}^{(\text{RIS})}$ and $\yv_{ij,t}^{(\text{MP})}$ represent the received signal at UE $j$ from the transmitting UE $i$ through the direct path, the reflected path through the RIS and the reflected signals from the SPs in the environment, respectively. Since the path degradation in mmWave is strong, we ignore the double-bounced paths (e.g., UE $i$-RIS-SP-UE $j$ channel) and only consider the NLoS components from the UE $i$-SP-UE $j$ channel \cite{cop-ris-2bs}. In \eqref{recived_signal}, $\nv_{j,t} \in \mathbb{C}^{N \times 1}$ is an additive white Gaussian noise (AWGN) at the $j$-th receiving UE where $\nv_{j,t}
\sim \mathcal{CN}(0, \sigma^2\mathbf{I}_N)$ with $\sigma^2$ being the noise variance, and  $\mathbf{I}_N$ is the identity matrix. The received LoS signal at the $j$-th UE from the $i$-th UE, at time $t$, can be defined as:
\begin{equation}
    \yv_{ij,t}^{(\text{LoS})} = \sqrt{E_i} \beta_{ij} \dv\left(\tau_{ij}\right) ,
    \label{recived_signal_los}
\end{equation}    
where $\beta_{ij}=\alpha_{ij}e^{j\rho_{ij}}$ is the complex channel gain, $\tau_{ij}$ is the transmission delay, and $\dv(\tau)$ is the delay steering vector defined as:
\begin{equation}
\dv(\tau)=
\begin{bmatrix}
1, \!\!& \!\! e^{-j2\pi\tau \Delta f}, & \!\!\!\! \ldots, \!\!\! & e^{-j2\pi\tau(N-1) \Delta_f} 
\end{bmatrix} ^\top,
\label{d_steer}
\end{equation}
where $\Delta_f$ is the subcarrier spacing and $(\cdot)^\top$ is the transpose operation. In \eqref{recived_signal_los}, $\tau_{ij} = d_{\text{U}_{i}\text{U}_{j}}/{c} + \Delta_{t_j} - \Delta_{t_i}$ where $c$ is the speed of light, $d_{\text{U}_{i}\text{U}_{j}} = {\left \lVert \pv_{\text{U}_i} - \pv_{\text{U}_j}  \right \rVert}_2$ represents the Euclidean distance between UEs $i$ and $j$, and $\Delta_{t_j}$ is the clock offset at UE $j$ with respect to a reference UE. Furthermore, the signal arriving from the uncontrolled multi-path propagation, $\yv_{ij,t}^{(\text{MP})}$, can be defined as:
\begin{equation}
    \yv_{ij,t}^{(\text{MP})} = \sqrt{E_i}\sum_{s=1}^{S_j} \beta_{ij,s} \dv\left(\tau_{ij,s}\right)  ,
    \label{recived_signal_mp}
\end{equation} 
where $S_j$ represents the total number of SPs seen by the receiving UE j. Here, $\beta_{ij,s}=\alpha_{ij,s}e^{j\rho_{ij,s}}$ and $\tau_{ij,s}$ are the complex channel gain and time delay of the uncontrolled path UE $i$ - SP $s$- UE $j$, respectively. Again, $\tau_{ij,s} = \left(d_{\text{U}_{i}\text{SP}_{j,s}} + d_{\text{U}_{j}\text{SP}_{j,s}} \right)/c + \Delta_{t_j} - \Delta_{t_i}$, where $d_{\text{U}_{i}\text{SP}_{j,s}} = {\left \lVert \pv_{\text{U}_i} - \pv_{\text{SP}_{j,s}}  \right \rVert}_2$ denotes the distance between UE $i$ and SP $s$ seen by the $j$-th receiving UE.

Finally, the signal arriving at UE $j$ from UE $i$ through the RIS can be modeled as:
\begin{equation}
    \yv_{ij,t}^{(\text{RIS})} = \sqrt{E_i} \beta_{ij,r} \dv\left(\tau_{ij,r}\right) \av^\top\left(\boldsymbol{\vartheta}_{j}\right) \Omegam_{i,t} \av\left(\boldsymbol{\vartheta}_{i}\right) ,
    \label{y_ris}
\end{equation}
where $\beta_{ij,r} = \alpha_{ij,r}e^{\rho_{ij,r}}$ is the complex channel gain of the path UE $i$-RIS-UE $j$, $\tau_{ij,r}$ is the transmission delay of the same path, and $\dv(\tau_{ij,r})$ is defined similar to \eqref{d_steer}. Here, $\tau_{ij,r} = \left( d_{\text{R}\text{U}_{i}} + d_{\text{R}\text{U}_{j}} \right) /c + \Delta_{t_j} - \Delta_{t_i}$, where $d_{\text{R}\text{U}_{i}} = {\left \lVert \pv_\text{R} - \pv_{\text{U}_i}  \right \rVert}_2$ represents the distance between UE $i$ and the RIS. The azimuth and elevation angles measured at the RIS center with respect to UE $i$ are represented by the vector $\boldsymbol{\vartheta}_{i}=[\theta_{az,i},\theta_{\text{el},i}]^\top$. The angles $\theta_{az,i}$ and $\theta_{el,i}$ are related to the position vector $\pv_{\text{U}_i}$ by: 
\begin{align}
    \theta_{az,i} &= \arctan\left(\frac{y_{\text{U}_i}}{x_{\text{U}_i}}\right), \label{eq:azi}\\
    \theta_{el,i} &= \arcsin\left(\frac{z_{\text{U}_i}}{\sqrt{x_{\text{U}_i}^2+y_{\text{U}_i}^2+z_{\text{U}_i}^2}} \right). \label{eq:ele}
\end{align} 
In \eqref{y_ris}, the vector $\av(\boldsymbol{\vartheta}_i) \in \mathbb{C}^{M \times 1}$ represents the RIS response vector with respect to the angle vector $\boldsymbol{\vartheta}_i$. The phase control matrix at the RIS at time $t$, when UE $i$ is the transmitter, is defined as:
\begin{equation}
    \begin{split}
        \Omegam_{i,t}&=\text{diag}(\omegav_{i,t}) \in \mathbb{C}^{M \times M}, \\
    \end{split}
    \label{ris_matrix}
\end{equation}
where
\begin{equation}
    \omegav_{i,t} = \begin{bmatrix}
        e^{j\omega_{i,t,1}},~ \ldots ,~ e^{ j\omega_{i,t,M}}
    \end{bmatrix}^\top \in \mathbb{C}^{M \times 1}.
    \label{ris_vector}
\end{equation}
 We can re-write \eqref{y_ris} as:
\begin{align}
\begin{split}
       \yv_{ij,t}^{(\text{RIS})} &= \sqrt{E_i} \beta_{ij,r} \dv\left(\tau_{ij,r}\right) \left(\av\left(\boldsymbol{\vartheta}_i\right)\odot \av\left(\boldsymbol{\vartheta}_j\right) \right)^\top \Omegam_{i,t} \boldsymbol{1}_M   \\
       &= \sqrt{E_i} \beta_{ij,r} \dv\left(\tau_{ij,r}\right) \bv^\top\left(\boldsymbol{\vartheta}_i,\boldsymbol{\vartheta}_j\right)  \omegav_{i,t}\\ 
       &= \sqrt{E_i} \beta_{ij,r} \dv\left(\tau_{ij,r}\right) \cv^\top\left(\boldsymbol{\gamma}_{ij}\right)  \omegav_{i,t},
       \label{recived_signal_ris}
       \end{split}
\end{align}
where $\odot$ represents the Hadamard product, $\bv\left(\boldsymbol{\vartheta}_i,\boldsymbol{\vartheta}_j\right) =\left(\av(\boldsymbol{\vartheta}_i)\odot \av(\boldsymbol{\vartheta}_j) \right)$, $\boldsymbol{1}_M$ is a column vector of length $M$ with all entries equal to $1$ and $\omegav_{i,t} = \Omegam_{i,t}\boldsymbol{1}_M$. The $m$-th element of the intermediate steering vector at the RIS, $\bv(\boldsymbol{\vartheta}_i,\boldsymbol{\vartheta}_j)$, can be computed as:
\begin{equation}\left[\bv\left(\boldsymbol{\vartheta}_i,\boldsymbol{\vartheta}_j\right)\right]_m = \left[\cv\left(\boldsymbol{\gamma}_{ij}\right)\right]_m =  e^{j\frac{2 \pi}{\lambda} \pv_{\text{R},m}^\top \boldsymbol{\gamma}_{ij}}, 
\label{eq_intermediate_steering_vector}
\end{equation}
where $\lambda$ is the wavelength of transmitted signal, $\boldsymbol{\gamma}_{ij}$ is the spatial frequency vector defined as  $\boldsymbol{\gamma}_{ij}=[\kappa_{ij},\xi_{ij},\zeta_{ij}]^\top$, and $\pv_{\text{R},m}$ represents the position of the $m$-th element of the RIS. Since the RIS is placed in the $yz$-plane, the first entry of $\pv_{\text{R},m}$ equals zero and hence $\kappa_{ij}$ is never utilized and will be ignored in the rest of the paper. To avoid angle ambiguity at the RIS, we set the RIS element spacing to be $\lambda /4$ similar to \cite{zero-ap-ris}. Hence, we can define the spatial frequencies as:
\begin{align}
\begin{split}
        \xi_{ij} &= \sin\left(\theta_{\text{az},i}\right)\cos\left(\theta_{\text{el},i}\right) + \sin\left(\theta_{\text{az},j}\right)\cos\left(\theta_{\text{el},j}\right), \\
    \zeta_{ij} &= \sin\left(\theta_{\text{el},i}\right) + \sin\left(\theta_{\text{el},j}\right). 
    \end{split}
    \label{intermediate_angle}
\end{align}
Consequently, only two spatial frequencies, $\xi_{ij}$ and $\zeta_{ij}$, can be estimated and thus we cannot estimate the angles $\theta_{\text{az},i}$, $\theta_{\text{el},i}$, $\theta_{\text{az},j}$ and $\theta_{\text{el},j}$ directly \cite{JrCUP}. From \eqref{intermediate_angle}, we observe that both $\xi_{ij}$ and $\zeta_{ij}$ are in the interval $[-2,~2]$. By substituting \eqref{eq:azi} and \eqref{eq:ele} in \eqref{intermediate_angle}, we get:
\begin{align}
    \begin{split}
        \xi_{ij}&= \frac{y_{\text{U}_i}}{\sqrt{x_{\text{U}_i}^2+y_{\text{U}_i}^2+z_{\text{U}_i}^2}}+\frac{y_{\text{U}_j}}{\sqrt{x_{\text{U}_j}^2+y_{\text{U}_j}^2+z_{\text{U}_j}^2}}, \\
        \zeta_{ij} &= \frac{z_{\text{U}_i}}{\sqrt{x_{\text{U}_i}^2+y_{\text{U}_i}^2+z_{\text{U}_i}^2}} + \frac{z_{\text{U}_j}}{\sqrt{x_{\text{U}_j}^2+y_{\text{U}_j}^2+z_{\text{U}_j}^2}}.
        \label{xi,zeta,location}
    \end{split}
\end{align}

In this paper, we only estimate the positions of the UEs and we are not estimating the positions of SPs as mapping of the environment is beyond the scope of the paper. Since the developed algorithm in the paper does not depend on multi-path signals, we simplify our analysis by dropping $\yv_{ij,t}^{(\text{MP})}$ from \eqref{recived_signal} and we evaluate the robustness of the proposed positioning algorithm to multi-path propagation in Section~\ref{sec:SP simulation}. Consequently, the received signal simplifies to:
\begin{equation}
   \yv_{ij,t} = \yv_{ij,t}^{(\text{LoS})}   
   + \yv_{ij,t}^{(\text{RIS})} +  \nv_{j,t}.\label{recived_signal_nosp}
\end{equation}
\subsection{Problem Statement}
Based on the channel model described above, we define an unknown channel parameter vector $\etav$:
\begin{equation}
    \etav = 
    \begin{bmatrix}
     \etav_{12}^\top,  \etav_{13}^\top, \ldots, 
     \etav_{1K}^\top, 
     \etav_{21}^\top,  \etav_{23}^\top, \ldots, \etav_{K, K-1}^\top
    \end{bmatrix} ^\top, 
\label{eta_vector}
\end{equation}
which contains $2 {\binom{K}{2}} = (K-1)K$ subvectors. Each subvector $ \etav_{ij} \in \mathbb{R}^{8}$ contains the channel parameters of the channel between the $i$-th UE (as the transmitter) and the $j$-th UE (as the receiver) given by: 
\begin{equation}
\begin{split}
     \etav_{ij} = 
     ~[\tau_{ij}, \tau_{ij,r}, \xi_{ij}, \zeta_{ij}, \alpha_{ij}, \rho_{ij}, \alpha_{ij,r}, \rho_{ij,r}]^\top.
    \end{split} \label{eq:unknown_channel_parameter_vector}
\end{equation}

We further define an unknown state vector $\sv \in \mathbb{R}^{4K - 1}$ (with channel gains removed) as:
\begin{equation}
    \sv = [\pv_{\text{U}_1}^\top, \ldots, \pv_{\text{U}_K}^\top, \Delta_{t_1}, \Delta_{t_2}, \ldots,\Delta_{t_{i-1}},\Delta_{t_{i+1}},\dots, \Delta_{t_K} ]^\top,
    \label{state_vector}
\end{equation}
where we take UE $i$ as a reference resulting in $\Delta_{t_i}=0$. By concatenating all the observed signal symbols $\yv = [\yv_{12}^\top, \ldots, \yv_{1K}^\top, \yv_{21}^\top, \ldots, \yv_{K, K-1}^\top]^\top$ with $\yv_{ij} = [\yv_{ij,1}^\top, \ldots, \yv_{ij,T}^\top]^\top$, the localization problem can be formulated as follows:
\begin{enumerate}
    \item Design the RIS configuration (i.e., the phase control matrix), which is dependent on whether or not prior information about the UEs' positions is available;
    \item Extract the channel parameters from the observed signals;
    \item Localization of all the UEs. 
\end{enumerate}
    
\subsection{Feasibility Analysis}
The complex channel gains will not contribute to estimating the locations of the UEs. By removing the nuisance parameters (i.e., channel gain parameters), we define a channel parameter vector $\bar\etav_{ij}\in \mathbb{R}^{4}$ as:
\begin{equation}
    \bar\etav_{ij} = [\tau_{ij}, \tau_{ij,r}, \xi_{ij}, \zeta_{ij}]^\top,
    \label{eq:channel_parameter_vector}
\end{equation}
and $\bar\etav$ can be obtained similarly based on~\eqref{eta_vector}.
Note that $\bar\etav_{ji}$ provides the same geometrical information as $\bar\etav_{ij}$, which can be treated as multiple measurements (more details in Section~\ref{sec:proposed_estimators}). Consequently, $\bar\etav$ (without the complex channel gains) can provide $4{\binom{K}{2}}=2K(K-1)$ geometrical equations to solve $4K -1$ state unknowns, which requires the minimum number of UEs to be $3$. For more challenging scenarios (e.g., blockage between UE-UE or UE-RIS), more UEs can enable localization by providing more geometrical information, which will not be discussed in this paper. 

\subsection{Cooperation Strategy}
As mentioned earlier, we assume that each UE takes the role of the transmitter once while the other UEs are in the receiving mode. To avoid interference, we set the UEs to be transmitters in a sequential order where UE $1$ is the first transmitter and then UE $2$, and so on. Each UE transmits $T$ OFDM signals. Consequently, there is a total of $KT$ transmissions per localization occasion. Since the proposed system model does not rely on APs or BSs, one of the available UEs can be used as a localization coordinator (i.e., a reference UE\footnote{Throughout the paper, we assume that the coordinating  UE is the reference UE for clock offset measurements and for the localization algorithm in Section~\ref{sec:localization_algorithm}.}) similar to \cite{RIS_magazine}. The coordinating UE is responsible for controlling the RIS and sharing the phase control matrices with other UEs. We assume that the UE transmission is synchronized with the phase profile changes at the RIS. To perform localization, each receiving UE estimates the channel parameters based on the received signals locally. Then, all UEs share the channel parameters information with the coordinating UE. Finally, the coordinating UE estimates its position and other UEs' positions, and distributes the position information back to all UEs.


\section{Derivation of Cramér-Rao Bound}
\label{sec:crlb}
In this section, we start by deriving the FIM for the unknown channel parameters $\etav$ defined in \eqref{eta_vector}. Then, the FIM for the UEs' positions can be derived by computing a transformation matrix that links the two matrices. Finally, the CRLB for the estimation error of the UEs' positions is derived, which serves as a benchmark for the estimators we develop in this paper. 

Let us define $\hat{\etav}$ as an unbiased estimate of $\etav$. Based on the Cramér–Rao theorem, the mean squared error (MSE) matrix of $\hat{\etav}$ is satisfied by the following inequality \cite{crb1,crb2}: 
\begin{equation}
    \mathbb{E}\{ \left( \hat{\etav} - \etav\right)\left( \hat{\etav} - \etav\right)^H     \} \succeq \mathbf{J}_{\etav}^{-1},
\end{equation}
where $\mathbf{J}_{\etav} $ is the FIM for $\etav$. The $(b,v)$-th entry for the FIM can be defined as:
\begin{equation}
    \left[ \mathbf{J}_{\etav} \right]_{b,v} = \mathbb{E} \left\{- \frac{\partial^2 \ln p(\yv;\etav )}{\partial \eta_b \partial \eta_v}     \right\},
    \label{J_bv}
\end{equation} 
where the $b$-th entry of $\etav$ is represented by $\eta_b$ and $p(\yv;\etav )$ represents the likelihood function of $\yv$ conditioned on $\etav$. Without loss of generality, similar to \cite{cop-ris-d2d}, we assume that the received signals $\yv_{ij}~\forall~i,~j,~i\neq j$ are independent. Hence, we can express $\ln p(\yv;\etav )$ as:
\begin{equation}
    \ln p(\yv;\etav ) = \sum_{i=1}^{K}\sum_{j\neq i}^{K}  \ln p(\yv_{ij};\etav ).
    \label{likelihood_sum}
\end{equation}
Based on \eqref{J_bv} and \eqref{likelihood_sum}, we can compute the FIM $\mathbf{J}_{\etav}$ as:
\begin{equation}
    \mathbf{J}_{\etav} = \sum_{i=1}^{K}\sum_{j\neq i}^{K}\mathbf{J}_{\etav}^{ij},
\end{equation}
where the $(b,v)$-th entry of $\mathbf{J}_{\etav}^{ij}$ can be obtained as follows:
\begin{equation}
    \left[ \mathbf{J}_{\etav} ^{ij} \right]_{b,v} = \frac{2}{\sigma^2}\sum_{t=1}^T \Re\left\{ \left(\frac{  \partial \muv_{ij,t}}{\partial \eta_b}  \right)^H \left(\frac{  \partial \muv_{ij,t}}{\partial \eta_v}  \right)    \right\},
    \label{FIM_general}
\end{equation} 
where $\Re\{\cdot\}$ represents the real part of a complex quantity, $(\cdot)^H$ is the conjugate transpose (Hermitian) operation, and
\begin{align}
\begin{split}
    \muv_{ij,t} &= \yv_{ij,t}^{(\text{LoS})}   
   + \yv_{ij,t}^{(\text{RIS})}  \\
   &= \sqrt{E_i} \beta_{ij} \dv\left(\tau_{ij}\right) + \sqrt{E_i} \beta_{ij,r} \dv\left(\tau_{ij,r}\right) \cv^\top\left(\boldsymbol{\gamma}_{ij}\right) \omegav_{i,t}.
   \end{split}
\end{align}
The partial derivatives needed for computing \eqref{FIM_general} are derived in Appendix~\ref{partial_signal_der}. 

Let us define $\mathbf{J}$ as the FIM associated with the state vector $\sv$ \eqref{state_vector} and the complex channel gains (i.e., $\alpha_{ij}, \alpha_{ij,r},\rho_{ij}$ and $\rho_{ij,r}~\forall~i\neq j$). $\mathbf{J}$ can be computed by applying the transformation matrix $\mathbf{T}$:
\begin{equation}
    \mathbf{J} = \mathbf{T} {\mathbf{J}}_{\etav} \mathbf{T}^T,
    \label{transform}
\end{equation}
 where $\mathbf{T} \in \mathbb{R}^{g\times f}$, for $g=4K-1+4K(K-1)$ and $f=8K(K-1)$, is defined as:
\begin{equation}
    \mathbf{T} = \frac{\partial {\etav}^T}{\partial \hv},
\end{equation}
where the number of columns in $\mathbf{T}$ (i.e., $f$) equals the length of the vector $\etav$ \eqref{eta_vector}, and $\hv$ can be written as:
\begin{equation}
   \hv = \begin{bmatrix}
        \sv^\top, \alpha_{ij},\ldots,\alpha_{ij,r},\ldots,\rho_{ij},\ldots,\rho_{ij,r},\ldots
    \end{bmatrix}^\top.
\end{equation}
The length of $\hv$ (i.e., $g$) equals the length of $\sv$ (i.e., $4K-1$) in addition to $4K(K-1)$ elements which are the total number of parameters related to the channel gains.
The partial derivatives of the channel parameters with respect to $\hv$ are derived in Appendix~\ref{partial_eta}. 
The first $4K-1$ rows (or columns) of $\mathbf{J}$ correspond to the UEs' positions and clock offsets while the next $4K(K-1)$ rows (or columns) correspond to the channel gains. 


Finally, the positioning error bound (PEB) of UE $i$ can be calculated as:
\begin{equation}
    \text{PEB}_i =\sqrt{ \text{tr}\left(\left[\mathbf{J}^{-1}\right]_{3i-2:3i,3i-2:3i}\right)}.
    \label{PEB_general}
\end{equation}
The clock offset error bound (CEB) at UE $i$ can be computed as:
\begin{equation}
    \text{CEB}_i =\sqrt{ \left(\left[\mathbf{J}^{-1}\right]_{3K+i,3K+i}\right)}.
\end{equation}

\section{RIS Profile Design}
\label{Sec:profile design}
In this section, we discuss orthogonal RIS profiles and how to tune the phase shifts of the RIS elements via different codebook designs. In addition, we discuss the power allocation between the UEs for a fixed codebook.
\subsection{Orthogonal RIS Profiles}
\label{sec:orthogonal_ris_profile}
To assist the channel parameters estimation in Section~\ref{sec:proposed_estimators}, we decompose the received signal into two components via orthogonal RIS profiles.
Similar to \cite{zero-ap-ris,siso-ris}, we design the RIS control matrices for half of the transmission time and represent them by $\Tilde{\Omegam}_{i,\Tilde{t}}$, where $\Tilde{t}=1,2,\ldots,\frac{T}{2}$ ($T$ is even). We discuss the choice of the RIS element coefficients in $\Tilde{\Omegam}_{i,\Tilde{t}}$ in the following subsections. Without loss of generality, let us define $\Omegam_{i,2\Tilde{t}-1}=\Tilde{\Omegam}_{i,\Tilde{t}}$ and $\Omegam_{i,2\Tilde{t}}=-\Tilde{\Omegam}_{i,\Tilde{t}}$. By subtracting and adding two consecutive received signals, we can obtain the following two quantities assuming that UE $i$ is the transmitter and UE $j$ is the receiver:
\begin{align}
\begin{split}
    \Tilde{\yv}_{ij,\Tilde{t}}^{(\text{RIS})} &= \frac{1}{2} \left( \yv_{ij,2\Tilde{t}-1} - \yv_{ij,2\Tilde{t}}  \right)   \\
    &= \sqrt{E_i} \beta_{ij,r} \dv\left(\tau_{ij,r}\right) \cv^\top\left(\boldsymbol{\gamma}_{ij}\right)  \Tilde{\Omegam}_{i,\Tilde{t}} \boldsymbol{1}_M + \Tilde{\nv}_{j,\Tilde{t}}\\
    &= \sqrt{E_i} \beta_{ij,r} \dv\left(\tau_{ij,r}\right) \cv^\top\left(\boldsymbol{\gamma}_{ij}\right)  \tilde \omegav_{i,\tilde t} + \Tilde{\nv}_{j,\Tilde{t}}, \label{rx-ris-separated}
    \end{split}\\
\begin{split}
    \Tilde{\yv}_{ij,\Tilde{t}}^{(\text{LoS})} &= \frac{1}{2} \left( \yv_{ij,2\Tilde{t}-1} + \yv_{ij,2\Tilde{t}}  \right)   \\
    &= \sqrt{E_i} \beta_{ij} \dv\left(\tau_{ij}\right)  +\Tilde{\nv}_{j,\Tilde{t}}, \label{rx-direct-separated}
    \end{split}
\end{align}
where \eqref{rx-ris-separated} follows from \eqref{recived_signal_ris} and \eqref{recived_signal_nosp}, and \eqref{rx-direct-separated} follows from \eqref{recived_signal_los} and \eqref{recived_signal_nosp}. The vector $\Tilde{\nv}_{j,\Tilde{t}} \in \mathbb{C}^{N \times 1}$ is an AWGN at UE $j$ where $\Tilde{\nv}_{j,\Tilde{t}}
\sim \mathcal{CN}(0, \frac{\sigma^2}{2}\mathbf{I}_N)$. We can see that we are able to decompose the received signal into two parts. The first part represents the signal arriving through the RIS path while the second part represents the LoS path. We can concatenate the received signals \eqref{rx-ris-separated} and \eqref{rx-direct-separated} over different transmissions into the matrices $\mathbf{Y}_{ij}^{(\text{RIS})}$ and $\mathbf{Y}_{ij}^{(\text{LoS})}$, respectively, which can be defined as:
\begin{align}
    \begin{split}
        \mathbf{Y}_{ij}^{(\text{RIS})} &= \begin{bmatrix}
            \Tilde{\yv}_{ij,1}^{(\text{RIS})},~ \Tilde{\yv}_{ij,2}^{(\text{RIS})},~ \ldots,~ \Tilde{\yv}_{ij,\frac{T}{2}}^{(\text{RIS})} 
        \end{bmatrix} ~\in \mathbb{C}^{N\times \frac{T}{2}},\\
        \mathbf{Y}_{ij}^{(\text{LoS})} &= \begin{bmatrix}
            \Tilde{\yv}_{ij,1}^{(\text{LoS})},~ \Tilde{\yv}_{ij,2}^{(\text{LoS})},~ \ldots,~ \Tilde{\yv}_{ij,\frac{T}{2}}^{(\text{LoS})} 
        \end{bmatrix} ~\in \mathbb{C}^{N\times \frac{T}{2}}.
    \end{split}
\end{align}
We concatenate the RIS control matrices over time for each transmitting UE $i$ into $\Tilde{\Omegam}_{i}\in \mathbb{C} ^{M \times \frac{T}{2}}$ which can be expressed as:
\begin{equation}
\begin{split}
    \Tilde{\Omegam}_{i} &=
    \begin{bmatrix}
    \Tilde{\Omegam}_{i,1}\boldsymbol{1}_M,~ \Tilde{\Omegam}_{i,2}\boldsymbol{1}_M,~ \ldots, ~ \Tilde{\Omegam}_{i,\frac{T}{2}}\boldsymbol{1}_M
\end{bmatrix}\\
&= \begin{bmatrix}
    \Tilde \omegav_{i,1}, ~\Tilde \omegav_{i,2}, ~ \ldots, ~\Tilde \omegav_{i,\frac{T}{2}}
\end{bmatrix}.
\end{split}
\label{omega_combined}
\end{equation}
In the following subsections, we provide two practical RIS profile codebooks and we introduce a power allocation optimization problem.
\subsection{Random Phase Codebook}
If no prior information on the UEs' positions is available, a natural way is to set the RIS coefficients as random phases. Since the RIS control matrix for the $i$-th transmitting UE and time slot $\Tilde{t}$ is defined as $\tilde \Omegam_{i,\tilde t}=\text{diag}(\tilde \omegav_{i,t}) \in \mathbb{C}^{M \times M}$ where $\tilde \omegav_{i,t} = [e^{j\tilde \omega_{i,\tilde t,1}}, \ldots , e^{ j\tilde \omega_{i,\tilde t,M}}]^\top $, we can choose the phase shift value at each RIS element from a uniform distribution over the interval $[0,2\pi)$ (i.e., $\tilde \omega_{i,\tilde t,m} \sim \mathcal{U}[0, 2\pi),~\forall~m,~i,~\tilde t$). 

\subsection{Directional Codebook}
\label{sec:directionl codebook}
If we have any prior information on the UEs' positions (e.g., from previous estimations), we can utilize this information to improve the positioning performance \cite{siso-ris}. A simple design is to reflect the incident signal on the RIS from the transmitting UE towards the direction of the receiving UE. However, both transmitting and receiving UEs' locations are unknown, and there are $K-1$ receiving UEs for each transmitting UE. To create the codebook, we choose the elements $\xi, \zeta$ in the directional vector $\boldsymbol{\gamma}$ based on some prior information and then the pahse shift at each RIS element can be set as the conjugate of~\eqref{eq_intermediate_steering_vector} as $e^{-j\frac{2 \pi}{\lambda} \pv_{\text{R},m}^\top \boldsymbol{\gamma}_{ij}}$, named as \textbf{directional codebook}.

Let us assume that the locations of the UEs follow a certain distribution based on the prior information (e.g., $\pv_{\text{U}_i}\sim \mathcal{N}(\tilde \pv_{\text{U}_i},\tilde \Sigmam_{\pv_{\text{U}_i}})$. Then, for each time slot $\tilde t$, we first sample the location of the transmitting UE $i$. Next, we sample uniformly the index of the receiving UE $j$ from $\{1,2,\ldots,K\}-\{i\}$ (since there are multiple receiving UEs) and then we sample the location of selected UE $j$ from the prior distribution. Based on the sampled locations of UEs $i$ and $j$, we can calculate $\boldsymbol{\gamma}_{ij}$ (i.e., $\xi_{ij}$ and $\zeta_{ij}$) using \eqref{xi,zeta,location}. The RIS profile can be written as:
\begin{equation}
    {[\tilde \omegav_{i,\tilde t}]}_m = e^{-j\frac{2 \pi}{\lambda} \pv_{\text{R},m}^\top \boldsymbol{\gamma}_{ij}}~\forall m,
\end{equation}
where $[\tilde \omegav_{i,\tilde t}]_m$ is the phase shift at the $m$-th RIS element for the $i$-th transmitting UE and time slot $\tilde t$. All RIS elements use the same value of $\boldsymbol{\gamma}_{ij}$ during a transmission time slot to maximize the signal energy towards a certain direction.

\subsection{Power Allocation Optimization}
The positioning performance can be improved by optimizing the power allocation between the UEs. Let us define $\boldsymbol{P}=[P_1,P_2,\dots,P_K]^\top$ as the power allocation vector where $P_i = N E_i$ is the transmitted power of UE $i$. We assume that there is a constraint on the total transmitted power in the system such that $\sum_{i=1}^K P_i=P_{\text{tot}}$ where $P_{\text{tot}}$ represents the total power transmitted in the system during each time slot. For a fixed RIS codebook, we minimize the average PEB of the involved UEs\footnote{Note that this is not the only way of formulating the power allocation problem. The objective function can also be expressed as minimizing the worst PEB.} by formulating the following optimization problem:
\begin{equation}
    \begin{split}
        \min_{\boldsymbol{P}}\ 
        & \frac{1}{K} \sum_{i=1}^K \text{PEB}_i
        \\
        \mathrm{s.t.\ \ } 
        & \sum_{i=1}^K P_i=P_{\text{tot}},\\ 
        &0 < P_i,
    \end{split}
    \label{eq_power_allocation_optimization}
\end{equation}
where $\text{PEB}_i$ is defined in \eqref{PEB_general}. The first constraint in \eqref{eq_power_allocation_optimization} ensures that the total transmitted power equals $P_{\text{tot}}$ and the second constraint ensures that each UE has a non-zero transmission power. The optimization problem \eqref{eq_power_allocation_optimization} is convex as discussed in \cite{power_allocation} and thus can be solved using any convex optimization toolbox. One limitation with the optimization problem \eqref{eq_power_allocation_optimization} is that the objective function (i.e., $\text{PEB}_i~\forall~i$) requires the positions of all the UEs, which are unknown. Assume we have the prior information of all the UEs following multi-variable Gaussian distributions similar to Section~\ref{sec:directionl codebook}, we can then calculate $\text{PEB}_i$ as $\text{PEB}_i(\tilde \pv_{\text{U}_1},\dots,\tilde \pv_{\text{U}_i},\dots,\tilde \pv_{\text{U}_K},\tilde \Omegam_1,\dots,\tilde \Omegam_K)$ where $\tilde \pv_{\text{U}_i}$ is the mean of the prior information of UE $i$ and $\tilde \Omegam_i$ is obtained from the directional codebook. This is not guaranteed to be optimal, but it is a practical solution for the RIS profile design.

\section{Proposed Estimators}
\label{sec:proposed_estimators}
In this section, we start by estimating the delays and spatial frequencies based on the received signals. Then, we develop a localization algorithm to estimate the UEs' positions.

\subsection{Delay Estimation}
\label{sec:delay_estimation}
We start by estimating the transmission delays of the signals reflected by the RIS (i.e., $\tau_{ij,r}~\forall~i,~j$). Coarse estimation of the delays can be computed by applying the inverse fast Fourier transform (IFFT) on the matrix $\mathbf{Y}_{ij}^{(\text{RIS})}$.
Let us define $\mathbf{F} \in \mathbb{C}^{N_f \times N} $ to be the IFFT matrix where $N_f$ represents the length of the IFFT vector, which is a design parameter. The $(l,h)$-th entry of the IFFT matrix can be computed as:
\begin{equation}
    [\mathbf{F}]_{l,h} = \frac{1}{N_f}e^{j2\pi lh/N_f}.
\end{equation}
The coarse estimates of the delays can be computed as:
\begin{equation}
    \bar{\tau}_{ij,r} = \frac{l_{ij,r}^*}{N_f \Delta f}  ~ \forall~i,~j,~i\neq j,
    \label{coarse_estimate}
\end{equation}
where $l_{ij,r}^*$ is an integer that can be computed as follows \cite{zero-ap-ris}:
\begin{equation}
    l_{ij,r}^* = \underset{l}{\arg\max} {\left \lVert \ev_l^\top \left(\mathbf{F} \mathbf{Y}_{ij}^{(\text{RIS})}\right)  \right \rVert}_2^2 ~ \forall~i,~j,~i\neq j,
\end{equation}
where $\ev_l$ is a column vector of length $N_f$ with all elements equal to zero except the $l$-th entry that has a value of $1$. In \eqref{coarse_estimate}, we assume that the delay estimate ($\bar{\tau}_{ij,r}$) is an integer multiple of the time resolution $\frac{1}{N_f \Delta f}$. To refine the delay estimate, we first define an intermediate variable $\delta_{ij,r} \in [0,\frac{1}{N_f \Delta f})$. Then, the optimal value of the intermediate variable can be calculated by \cite{siso-ris}: 
\begin{equation}
    \bar{\delta}_{ij,r} =  \underset{\delta_{ij,r}}{\arg\max} {\left \lVert \ev_{l_{ij,r}^*}^\top \mathbf{F} \left( \mathbf{Y}_{ij}^{(\text{RIS})} \odot \mathbf{D}(\delta_{ij,r}) \right)  \right \rVert}_2^2 ~ \forall~i,~j,~i\neq j,
    \label{eq:delay_optim}
\end{equation}
where $\mathbf{D}(\delta_{ij,r})$ is defined as:
\begin{equation}
    \mathbf{D}(\delta_{ij,r}) =\underbrace{ \begin{bmatrix}
        \dv(\delta_{ij,r}),~ \ldots,~ \dv(\delta_{ij,r})
    \end{bmatrix}}_{\frac{T}{2}~\text{times}} ~\in~\mathbb{C}^{N \times \frac{T}{2}}.
\end{equation} 
The above optimization problem \eqref{eq:delay_optim} can be solved using a quasi-Newton method by setting the starting point to $0$.
Next, the refined delay estimate can be obtained as:
\begin{equation}
    \Tilde{\tau}_{ij,r} = \bar{\tau}_{ij,r} - \bar{\delta}_{ij,r} ~ \forall~i,~j,~i\neq j.
    \label{refine_estimate}
\end{equation}

We observe that the delays $\tau_{ij,r}$ and $\tau_{ji,r}$ provide the same geometrical information. Due to the channel reciprocity and coherence assumption, we can simply average the two estimates as:
\begin{equation}
    \hat{\tau}_{ij,r} = \frac{\Tilde{\tau}_{ij,r} + \Tilde{\tau}_{ji,r}}{2}.
    \label{final_delay_estimate}
\end{equation}
Since $\tau_{ij,r} = \left( d_{\text{R}\text{U}_{i}} + d_{\text{R}\text{U}_{j}} \right) /c + \Delta_{t_j} - \Delta_{t_i}$ and $\tau_{ji,r} = \left( d_{\text{R}\text{U}_{i}} + d_{\text{R}\text{U}_{j}} \right) /c + \Delta_{t_i} - \Delta_{t_j}$, the clock offsets cancel out when we take the average of the two estimates. Thus, we do not need to estimate the clock offsets anymore (we can think of this as a two-way communication).

Similarly, we can estimate the delay of the LoS path (i.e., $\hat{\tau}_{ij}~\forall~i,~j,~i\neq j$) using \eqref{coarse_estimate}--\eqref{final_delay_estimate} by replacing $\mathbf{Y}_{ij}^{(\text{RIS})}$, $\bar{\tau}_{ij,r}$, $\Tilde{\tau}_{ij,r}$, $\hat{\tau}_{ij,r}$, $l_{ij,r}^*$ and $\bar{\delta}_{ij,r}$,  with $\mathbf{Y}_{ij}^{(\text{LoS})}$, $\bar{\tau}_{ij}$, $\Tilde{\tau}_{ij}$, $\hat{\tau}_{ij}$, $l_{ij}^*$ and $\bar{\delta}_{ij}$, respectively. 
\subsection{Spatial Frequency Estimation}
\label{sec:spatial_freq_estimation}
In this section, we estimate the spatial frequencies $\xi_{ij}$ and $\zeta_{ij}$. We start by eliminating the effect of the transmission delay on the received signal through the RIS path by computing:
\begin{align}
\begin{split}
    \mathbf{Y}_{ij}^{(\text{a})} &= \mathbf{Y}_{ij}^{(\text{RIS})}\odot \mathbf{D}\left(-\tilde{\tau}_{ij,r}\right).
    \label{y_angle}
\end{split}
\end{align}
We then take the sum of all rows of the matrix in \eqref{y_angle} (i.e., taking the sum over all subcarriers) and we get the following quantity:
\begin{align}
\begin{split}
 \Tilde{\yv}_{ij}^{(\text{r})}&=\mathbf{Y}_{ij}^{{(\text{a})}^\top} \boldsymbol{1}_N \\
    &= N \sqrt{E_i} \beta_{ij,r} \Tilde{\Omegam}_{i}^\top \cv\left(\boldsymbol{\gamma}_{ij}\right) + \Tilde{\qv},
    \label{y_r}
    \end{split}
\end{align}
where $\Tilde{\qv}\in\mathbb{C}^{\frac{T}{2}\times 1}$ represents the noise component and $\Tilde{\Omegam}_{i}$ is defined in \eqref{omega_combined}. 
From \eqref{y_r} and by ignoring the noise component, we can evaluate $\beta_{ij,r}$ as a function of $\boldsymbol{\gamma}_{ij}$ and thus we obtain:
\begin{equation}
    \Tilde{\beta}_{ij,r}\left(\boldsymbol{\gamma}_{ij} \right) = \frac{ \cv^H \left(\boldsymbol{\gamma}_{ij} \right) \Tilde{\Omegam}_{i}^* \Tilde{\yv}_{ij}^{(\text{r})} } {N \sqrt{E_i} {\left \lVert \cv^H \left(\boldsymbol{\gamma}_{ij} \right) \Tilde{\Omegam}_{i}^* \right \rVert}_2^2},
\end{equation}
where $(\cdot)^*$ is the conjugate operation. Finally, we can estimate $\xi_{ij}$ and $\zeta_{ij}$ by solving the following optimization problem:
\begin{align}
\begin{split}
    \Tilde{\xi}_{ij},~\Tilde{\zeta}_{ij} &= \underset{\xi_{ij},~\zeta_{ij}}{\arg\min} {\left \lVert \Tilde{\yv}_{ij}^{(\text{r})} -  N \sqrt{E_i} \Tilde{\beta}_{ij,r}\left(\boldsymbol{\gamma}_{ij} \right) \Tilde{\Omegam}_{i}^\top  \cv\left(\boldsymbol{\gamma}_{ij}\right)  \right \rVert}_2^2 \\ 
    &\textrm{s.t.} \quad -2 \leq \xi_{ij},~ \zeta_{ij} \leq 2, 
    \label{spatial_optimization}
    \end{split},
\end{align}
where the constraints follow from \eqref{intermediate_angle}. Coarse estimates of $\xi_{ij}$ and $\zeta_{ij}$ can be obtained by solving the above optimization problem via searching over a 2D grid, whose area can also be reduced with prior UE position information. Once coarse estimates are computed, we can refine the estimates by utilizing a quasi-Newton method and using the coarse estimates as the starting point. By looking at \eqref{intermediate_angle} and due to the reciprocity of the channel, we observe that $\xi_{ij}$ and $\xi_{ji}$ are equal. This equality applies to $\zeta_{ij}$ and $\zeta_{ji}$ as well. Hence, we can average the two estimates as:
\begin{align}
\begin{split}
    \hat{\xi}_{ij} &= \frac{\Tilde{\xi}_{ij} + \Tilde{\xi}_{ji}}{2},\\
    \hat{\zeta}_{ij} &= \frac{\Tilde{\zeta}_{ij} + \Tilde{\zeta}_{ji}}{2}.
    \label{final_angle_estimate}
    \end{split}
\end{align}
\comm{
\begin{algorithm}[!t]
 \caption{Spatial Frequency Estimation }
 \label{spatial-algorithm}
 \begin{algorithmic}[1]
 \renewcommand{\algorithmicrequire}{\textbf{Input:}}
 \renewcommand{\algorithmicensure}{\textbf{Output:}}
 \REQUIRE $\mathbf{Y}_{ij}^{(\text{RIS})}$, $\Tilde{\tau}_{ij,r}~\forall~i,~j,~i\neq j$
 \ENSURE  $\hat{\xi}_{ij}$, $\hat{\zeta}_{ij}~\forall~i,~j,~i\neq j$
  \FOR{$i = 1$ to $K$}
    \FOR{$j = 1$ to $K$}
        \IF {($i \ne j$)}
            \STATE $\mathbf{Y}_{ij}^{(\text{a})} \leftarrow \mathbf{Y}_{ij}^{(\text{RIS})}\odot \mathbf{D}\left(-\tilde{\tau}_{ij,r}\right)$
            \STATE $\Tilde{\yv}_{ij}^{(\text{r})} \leftarrow \mathbf{Y}_{ij}^{{(\text{a})}^\top} \boldsymbol{1}_N$
            \STATE $\Tilde{\xi}_{ij},~ \tilde{\zeta}_{ij} \leftarrow$ solve \eqref{spatial_optimization} by 2D grid search
            \STATE $\Tilde{\xi}_{ij},~ \tilde{\zeta}_{ij} \leftarrow$ refine the estimates using a quasi-Newton method by using the grid search results as starting point 
        \ENDIF
    \ENDFOR
  \ENDFOR
  \FOR{$i=1$ to $K$}
    \FOR{$j=i+1$ to $K$}
        \STATE $\hat{\xi}_{ij},~ \hat{\xi}_{ji} \leftarrow (\Tilde{\xi}_{ij}+\Tilde{\xi}_{ji})/2 $
        \STATE $\hat{\zeta}_{ij,r},~ \hat{\zeta}_{ji,r} \leftarrow (\Tilde{\zeta}_{ij,r}+\Tilde{\zeta}_{ji,r})/2 $
    \ENDFOR
  \ENDFOR
 \RETURN $\hat{\xi}_{ij}$, $\hat{\zeta}_{ij,r}~\forall~i,~j,~i\neq j$ 
 \end{algorithmic} 
 \end{algorithm}
 }
To summarize, for every transmitting UE $i$ and receiving UE $j$, we estimate four channel parameters (two time delays and two spatial frequencies). 
\subsection{Localization Algorithm}
\label{sec:localization_algorithm}
To estimate the UEs' positions, we first need to compute the angles at the RIS center with respect to each UE using the estimated spatial frequencies. Then, the angle and time delay estimates are used to localize the UEs. 
\subsubsection{Angle Estimation}
The estimated spatial frequencies can be concatenated as:
\begin{equation}
    \begin{split}
        \hat{\xiv} &= \begin{bmatrix}
            \hat{\xi}_{12}, \ldots, \hat{\xi}_{ij}, \ldots\end{bmatrix}^\top, 
            \\      
            \hat{\zetav} &= \begin{bmatrix}
            \hat{\zeta}_{12}, \ldots, \hat{\zeta}_{ij}, \ldots \end{bmatrix}^\top, 
    \end{split}
\end{equation}
where each of the vectors $\hat{\xiv}$ and $\hat{\zetav}$ has $K(K-1)/2$ elements. For example, if $\hat{\xiv}$ contains $\hat{\xi}_{12}$, it will not contain $\hat{\xi}_{21}$ since both $\hat{\xi}_{12}$ and $\hat{\xi}_{21}$ are equal due to the averaging in \eqref{final_angle_estimate}. We can re-write \eqref{intermediate_angle} as:
\begin{align}
\begin{split}
        \xi_{ij} &= w_{1,i} + w_{1,j}, \\
    \zeta_{ij} &= w_{2,i} + w_{2,j},
    \end{split}
    \label{spatial_frequncy_sum_w}
\end{align}
where $w_{1,i} = \sin\left(\theta_{\text{az},i}\right)\cos\left(\theta_{\text{el},i}\right)$
and $w_{2,i} = \sin\left(\theta_{\text{el},i}\right)$ which can be concatenated into the following two vectors:
\begin{align}
    \begin{split}
        \wv_1 &= \begin{bmatrix}
            w_{1,1},w_{1,2},\ldots,w_{1,K}
            \end{bmatrix}^\top, \\
         \wv_2 &= \begin{bmatrix}
            w_{2,1},w_{2,2},\ldots,w_{2,K}
            \end{bmatrix}^\top.   
    \end{split}
\end{align}
Based on \eqref{spatial_frequncy_sum_w}, we can relate $\xiv$ and $\wv_1$, and $\zetav$ and $\wv_2$ as follows: 
\begin{align}
    \begin{split}
        \xiv &= \Gm \wv_1, \\
        \zetav &= \Gm \wv_2, \\
    \end{split}
\end{align}
where $\Gm\in\mathbb{R}^{\frac{(K-1)K}{2} \times K}$, with the $k$-th row corresponding to the $k$-th element in $\xiv$ and $\zetav$. For example, for the $k$-th element $\xi_{ij}$, the entries $i$ and $j$ in the $k$-th row of the matrix $\Gm$ are set as 1 with all other entries as zeros. Based on the estimated spatial frequency vectors $\hat{\xiv}$ and $\hat{\zetav}$, we can estimate $\hat{\wv}_1$ and $\hat{\wv}_2$ using the least squares solution, which requires $K\geq 3$, as \cite{linear_algebra_book}:
\begin{align}
    \begin{split}
        \hat{\wv}_1 &= \left(\Gm^\top \Gm \right)^{-1} \Gm^\top \hat{\xiv},\\
        \hat{\wv}_2 &= \left(\Gm^\top \Gm \right)^{-1} \Gm^\top \hat{\zetav}.
    \end{split}
\end{align}
Consequently, the angle information of UE $i$ can be estimated as: 
\begin{align}
    \hat \theta_{\text{az}, i} & = \arcsin{(\hat w_{1,i}/\sqrt{1-\hat w_{2,i}^2})},\\
    \hat \theta_{\text{el}, i} & = \arcsin{(\hat w_{2,i})},
\end{align}
which can be concatenated into the vector $\hat{\varthetav}_i=[\hat{\theta}_{\text{az},i}, \hat{\theta}_{\text{el},i}]^\top$. 
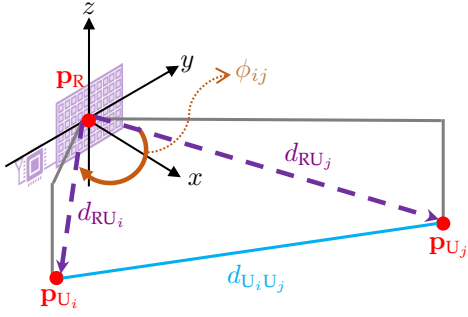
\begin{figure}
    \centering
    \begin{tikzpicture}
    \node (image) [anchor=south west]{\includegraphics[width=.75\linewidth]{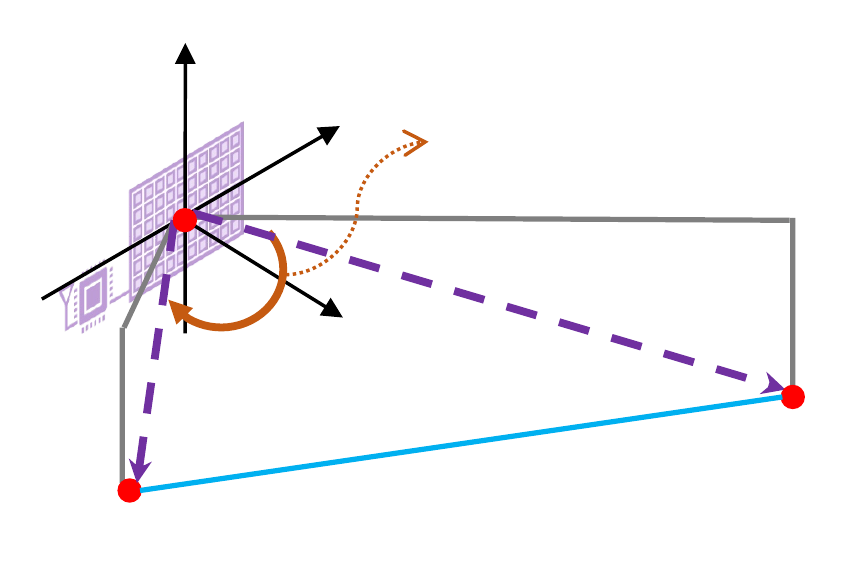}};
    \gettikzxy{(image.north east)}{\ix}{\iy};
    \node at (0.43*\ix,0.48*\iy)[rotate=0,anchor=north]{{$x$}};
    \node at (0.415*\ix,0.82*\iy)[rotate=0,anchor=north]{{$y$}};
    \node at (0.225*\ix,0.97*\iy)[rotate=0,anchor=north]{{$z$}};
    \node at (0.19*\ix,0.775*\iy)[rotate=0,anchor=north]{{\red{$\pv_\text{R}$}}};
    \node at (0.17*\ix,0.15*\iy)[rotate=0,anchor=north]{{\red{$\pv_{\text{U}_i}$}}};
    \node at (0.92*\ix,0.3*\iy)[rotate=0,anchor=north]{{\red{$\pv_{\text{U}_j}$}}};
    \node at (0.54*\ix,0.81*\iy)[rotate=0,anchor=north]{{\brown{$\phi_{ij}$}}};
    \node at (0.55*\ix,0.22*\iy)[rotate=0,anchor=north]{{\cyan{$d_{\text{U}_i\text{U}_j}$}}};
    \node at (0.25*\ix,0.4*\iy)[rotate=0,anchor=north]{{\violet{$d_{\text{RU}_i}$}}};
    \node at (0.65*\ix,0.57*\iy)[rotate=0,anchor=north]{{\violet{$d_{\text{RU}_j}$}}};
    \end{tikzpicture}
    \caption{Every two UEs $i$ and $j$ along with the RIS center forms a triangle. $\phi_{ij}$ represents the angle between the two directional vectors pointing towards each of UE $i$ and UE $j$.}
    \label{Fig:triangle}\vspace{-5mm}
\end{figure}
\subsubsection{Coarse Estimation of UEs' Positions}
Based on the angle and time delay estimates, we design a 1D search algorithm for UE positioning along the direction of the reference UE (i.e., the coordinating UE). By taking UE $i$ as our reference, for a given candidate distance between UE $i$ and the RIS (denoted as $\check d_{\text{RU}_i}$), a candidate location for the UE will be:
\begin{equation}
    \check \pv_{\text{U}_i} = \check d_{\text{RU}_i} \tv(\hat \varthetav_i ), 
\end{equation}
where $\tv(\hat\varthetav_i )$ is the direction vector of UE $i$ with respect to the RIS center defined as:
\begin{equation}
    \tv(\hat \varthetav_i ) = \begin{bmatrix}
        \cos(\hat \theta_{\text{az},i})\cos(\hat \theta_{\text{el},i})\\ \sin(\hat \theta_{\text{az},i})\cos(\hat\theta_{\text{el},i})\\
        \sin(\hat\theta_{\text{el},i})
    \end{bmatrix}.
\end{equation}
Each pair of UEs (e.g., $i$ and $j$), along with the RIS center, forms a triangle as shown in Figure~\ref{Fig:triangle}. Based on the law of cosines, the relationship between the $i$-th UE, the $j$-th UE ($j\ne i$), and the RIS is given by \cite{law_of_cosine}:
\begin{equation}
    d_{\text{R}\text{U}_{i}}^2 + d_{\text{R}\text{U}_{j}}^2 - 2d_{\text{R}\text{U}_{i}}d_{\text{R}\text{U}_{j}} \cos(\phi_{ij}) = d_{\text{U}_i\text{U}_j}^2,
    \label{law_of_cosine_with_phi}
\end{equation}
where $d_{\text{R}\text{U}_{i}}$, $d_{\text{R}\text{U}_{j}}$ and $d_{\text{U}_{i}\text{U}_{j}}$ denote the distances between UE $i$ and the RIS, between UE $j$ and the RIS, and between UE $i$ and UE $j$, respectively. The angle between the directional vectors $\tv(\varthetav_i)$ and $\tv(\varthetav_j)$ is represented by $\phi_{ij}$. Since $\cos(\phi_{ij})=\tv^\top(\varthetav_i) \tv(\varthetav_j)$, we re-write \eqref{law_of_cosine_with_phi} as:
\begin{equation}
    d_{\text{R}\text{U}_{i}}^2 + d_{\text{R}\text{U}_{j}}^2 - 2d_{\text{R}\text{U}_{i}}d_{\text{R}\text{U}_{j}} \tv^\top(\varthetav_i) \tv(\varthetav_j) = d_{\text{U}_i\text{U}_j}^2.
    \label{law_of_cosine}
\end{equation}
In \eqref{law_of_cosine_with_phi} and \eqref{law_of_cosine}, $d_{\text{U}_i\text{U}_j}$ can be calculated from the time delay of the LoS path as $c\tau_{ij}$. However, $d_{\text{RU}_i}$ and $d_{\text{RU}_j}$ can not be calculated directly from the time delay estimates. Using the time delay of the RIS path, we can calculate $d_{\text{RU}_i}+d_{\text{RU}_j}$ as $c\tau_{ij,r}$. In \eqref{law_of_cosine}, we can show that $d_{\text{U}_i\text{U}_j}^2$ is equivalent to $(c\tau_{ij, r} - c\tau_{ij} - d_{\text{R}\text{U}_{i}} - d_{\text{R}\text{U}_{j}})^2$. For a given $\check d_{\text{RU}_i}$ and by replacing $d_{\text{U}_i\text{U}_j}^2$ with its equivalent expression, we can solve \eqref{law_of_cosine} for $\check d_{\text{RU}_j}$ as: 
\begin{equation}
     \check d_{\text{R}\text{U}_{j}} = \frac{
    2(c \hat \tau_{ij,r} - c \hat \tau_{ij}) \check d_{\text{R}\text{U}_{i}} - (c \hat\tau_{ij,r} - c \hat \tau_{ij})^2}{2 \check d_{\text{R}\text{U}_{i}}(1+\tv^\top(\hat \varthetav_i) \tv(\hat \varthetav_j)) + 2(c \hat \tau_{ij} - c \hat \tau_{ij,r})
    }.
    \label{eq:non_ref_UE_distance}
\end{equation}
Then, the candidate position of all the other UEs can be obtained as: 
\begin{equation}
     \check \pv_{\text{U}_j}= \check d_{\text{R}\text{U}_{j}}\tv ( \hat \varthetav_j), \ \  j\ne i.
    \label{eq:non_ref_UE_position}
\end{equation}
The final estimation of the distance between the $i$-th UE and the RIS can be calculated by formulating a cost function as:
\begin{equation}
    d_{\text{R}\text{U}_{i}}^* = \argmin_{ d_{\text{R}\text{U}_{i}}}\underset{j\ne k \ne i} {\sum_{j,k}}
    \left( \check\psi_{jk} - c(\hat\tau_{jk,r} - \hat\tau_{jk}) \right)^2. 
    \label{coarse_location_cost_function}
\end{equation}
where $\check \psi_{jk} = \lVert  \check \pv_{\text{U}_j} - \pv_{\text{R}} \rVert + \lVert \check \pv_{\text{U}_k} - \pv_{\text{R}} \rVert - \lVert \check \pv_{\text{U}_j} - \check \pv_{\text{U}_k} \rVert$.
The cost function \eqref{coarse_location_cost_function} calculates the mismatch between the distance measurements based on the location estimates and the distance measurements based on the time delay estimates for each pair of UEs not including the reference UE $i$. Once the optimal $d_{\text{R}\text{U}_{i}}^*$ is found, the rest of the state parameters can be calculated based on~\eqref{eq:non_ref_UE_distance} and~\eqref{eq:non_ref_UE_position}. The 1D search along the direction of the reference UE $i$ provides coarse estimates of the UEs' positions which will be refined in the next section. We can do the 1D search using distance values given by the vector $\dv_{\text{search}}$ which has a total of $L$ elements. The steps of the coarse estimation of UEs' positions are summarized in Algorithm~\ref{coarse-position-algorithm}. We can then concatenate the outputs of Algorithm~\ref{coarse-position-algorithm} into the vector $\bar \sv$ defined as\footnote{Note that the clock offsets are not part of the state vector anymore since they cancel out when we average the delay estimates and they do not affect the positioning algorithm. }: 
\begin{equation}
    \bar \sv = \begin{bmatrix}
        \bar \pv_{\text{U}_1}^\top, \ldots, \bar \pv_{\text{U}_K}^\top
    \end{bmatrix}^\top.
    \label{coarse_estimate_state}
\end{equation}
\begin{algorithm}[t]
 \caption{Coarse Estimation of UEs' positions }
 \label{coarse-position-algorithm}
 \begin{algorithmic}[1]
 \renewcommand{\algorithmicrequire}{\textbf{Input:}}
 \renewcommand{\algorithmicensure}{\textbf{Output:}}
 \REQUIRE $\hat\varthetav_j, \hat \tau_{jk}, \hat \tau_{jk,r}$$~\forall~j,~k,~j\neq k$, $i$ (index of reference UE) and $\dv_{\text{search}}$
 \ENSURE  $\bar \pv_{\text{U}_j}~\forall~j$
  \STATE $d_{\text{R}\text{U}_{i}}^* \leftarrow  \displaystyle \argmin_{ d_{\text{R}\text{U}_{i}}\in \dv_{\text{search}}}\underset{j\ne k \ne i} {\sum_{j,k}} \left( \check\psi_{jk} - c(\hat\tau_{jk,r} - \hat\tau_{jk}) \right)^2$
  \STATE $\bar \pv_{\text{U}_i} \leftarrow  d_{\text{R}\text{U}_{i}}^*\tv(\hat \varthetav_i ) $
  \FOR{$j = 1$ to $K$}
        \IF {($j \ne i$)}
            \STATE $\bar d_{\text{RU}_j} \leftarrow$ calculate using \eqref{eq:non_ref_UE_distance} and $d_{\text{R}\text{U}_{i}}^*$
            \STATE $\bar \pv_{\text{U}_j} \leftarrow \bar d_{\text{RU}_j} \tv(\hat \varthetav_j ) $
        \ENDIF
    \ENDFOR
 \RETURN $\bar \pv_{\text{U}_j}~\forall~j$
 \end{algorithmic} 
 \end{algorithm}


\subsubsection{Maximum Likelihood Estimation}
With the coarse estimation of the state parameters (i.e., $\bar \sv$), refinement procedures can be performed by solving an \ac{mle} problem. The objective function can be formulated as
\begin{equation}
    \hat \sv = \argmin_{\sv}\ \left(\hat \etav - \etav(\sv)\right)^\top\Sigmam_{\etav}^{-1}\left(\hat \etav - \etav(\sv)\right),
    \label{eq_mle_position_estimation}
\end{equation}
where $\hat \etav$ is the estimated nuisance-free channel parameters (i.e., two delay estimates and two spatial frequency estimates for each pair of UEs $i$ and $j$), $\etav(\sv)$ is the mapping from the state parameters to the channel parameters and $\Sigmam_{\etav}$ is the covariance matrix of the estimated channel parameters. If the covariance matrix $\Sigmam_{\etav}$ is not available, we replace it with the identity matrix and thus \eqref{eq_mle_position_estimation} becomes a least squares optimization problem. In general, the optimization problem in~\eqref{eq_mle_position_estimation} can be solved by standard multi-variable optimization methods, such as the trust-region method \cite{manopt} and using the coarse estimate of the state parameters (i.e., $\bar \sv$ in \eqref{coarse_estimate_state}) as an initial point.

\section{Simulation Results}
\label{sec:simulation}
In this section, we evaluate the performance of the proposed RIS CP algorithm, benchmarked by the derived CRLB. We start by describing the simulation environment and then we discuss the numerical results generated over different system parameters.

\subsection{Simulation Setup}
\label{sec:simulation_setup}
We consider the scenario shown in Figure~\ref{scenario} as our simulation environment. We assume that there are three UEs which need to be localized. There is also one RIS placed along the $yz$-plane. The center of the RIS (i.e., $\pv_{\text{R}}$) is located at the origin $[0,0,0]^\top$ m. The dimensions of the simulation environment are $[0,7]$ m, $[-3.5, 3.5]$ m and $[-2,2]$ m on the $x$-axis, $y$-axis and $z$-axis, respectively. The three UEs: UE $1$, UE $2$ and UE $3$ are located at $\pv_{\text{UE}_1}=[4, 3, -1]^\top$ m, $\pv_{\text{UE}_2}=[4.5, 1, -0.5]^\top$ m and $\pv_{\text{UE}_3}=[5, -3, -1]^\top$ m, respectively. The distances between the RIS and each of the UEs are $d_{\text{RU}_1} \approx 5$ m, $d_{\text{RU}_2} \approx 4.6$ m and $d_{\text{RU}_3} \approx 6$ m. In addition, the distances between each pair of UEs are $d_{\text{U}_1\text{U}_2}\approx 2$ m, $d_{\text{U}_1\text{U}_3}\approx 6$ m and $d_{\text{U}_2\text{U}_3}\approx 4$ m. The communication channel parameters are summarized in Table~\ref{table:simulation_parameters} which are consistent with the simulation parameters in \cite{siso-ris, ris-loc-siso}. The total transmission power in the system $P_{\text{tot}}$ is set to $600$ mW or equivalently $27.8$ dBm. We assume that $P_{\text{tot}}$ is evenly distributed between the three UEs (i.e., $P_i=200$ mW or $23$ dBm $\forall~i$). The number of SPs is set to zero (i.e., $S_j=0~\forall j$) unless otherwise stated. The channel gain of each path in the simulations is calculated as \cite{wymeersch2022radio, wymeersch2022radio2, gain_ref} :  
\begin{align}
    \begin{split} \label{gain_1}
    \beta_{ij} &= \alpha_{ij} e^{j\rho_{ij}} = \frac{\lambda}{4\pi d_{\text{U}_i\text{U}_j}}e^{-j\frac{2\pi}{\lambda}d_{\text{U}_i\text{U}_j} },
    \end{split}\\
    \begin{split} \label{gain_2}
    \beta_{ij,r} &= \alpha_{ij,r} e^{j\rho_{ij,r}} = \frac{\lambda^2}{16\pi^2 d_{\text{R}\text{U}_i}d_{\text{R}\text{U}_j}}e^{-j\frac{2\pi}{\lambda}\left(d_{\text{R}\text{U}_i} + d_{\text{R}\text{U}_j}  \right) },
    \end{split}\\
    \begin{split} \label{gain_3}
    \beta_{ij,s} &= \alpha_{ij,s} e^{j\rho_{ij,s}}\\ &= \sqrt{\frac{\sigma^2_{\text{rcs}}}{4\pi}}\frac{\lambda}{4\pi d_{\text{U}_i\text{SP}_{j,s}}d_{\text{U}_j\text{SP}_{j,s}}}e^{-j\frac{2\pi}{\lambda}\left(d_{\text{U}_i\text{SP}_{j,s}} + d_{\text{U}_j\text{SP}_{j,s}}  \right) },
    \end{split}
\end{align}
where \eqref{gain_1}--\eqref{gain_3} are the channel gains for the LoS channel, the RIS channel, and the multi-path channel, respectively. The radar cross-section (RCS) $\sigma^2_{\text{rcs}}$ indicates the reflection capability of the NLoS path in terms of $\text{m}^2$. Note that $\beta_{i,j}$, $\beta_{ij,r}$, $\beta_{ij,s}$ are unknowns to be estimated. Finally, we use the root-mean-squared-error (RMSE) as a metric for evaluation.
\begin{table}[!t]
\caption{Simulation Parameters}
\begin{tabular}{ll}
\hline \hline
Parameter                      & Value                                  \\ \midrule \midrule
Carrier Frequency              & $28$ GHz                               \\
Wavelength                     & $\lambda=1$ cm                         \\
RIS array size                 & $M=11\times 11$ elements       \\
Speed of light                 & $c = 3 \times 10^8$ m/s                \\
Number of UEs                  & $K=3$                                  \\
Number of transmissions per UE & $T=40$                                 \\
Total number of transmission   & $KT=120$                               \\
RIS element spacing            & $\lambda/4=0.25$ cm                    \\
Number of subcarriers          & $N=3000$                               \\
subcarrier spacing             & $\Delta_f=120$ kHz                     \\
Symbol duration                & $8.33$ $\mu$s                          \\
Noise figure                   & $n_f = 8$ dB                           \\
Noise Power Spectral Density   & $N_0=-174$ dBm/Hz                      \\
Noise variance                 & $\sigma^2=n_f N_0 \Delta_f=-115.2$ dBm \\
Transmission power per UE             & $P_i=NE_i=23$ dBm                          \\
Total transmission power            &$P_{\text{tot}}=27.8$ dBm \\
Length of IFFT                 & $N_f = 10N$  \\            \hline \hline            
\end{tabular} \vspace{-5mm}
\label{table:simulation_parameters}
\end{table}

\subsection{Effect of Transmission Power on Positioning Performance }
In this section, we evaluate the performance of the proposed estimators by comparing their performance to the CRLB as a function of the transmission power. We vary the transmission power from $6$ dBm to $30$ dBm for the three UEs. The RIS control matrix is designed using the random phase codebook. The lower bounds are computed using one realization of the RIS profile. We average the results of the estimators over $200$ noise realizations.  

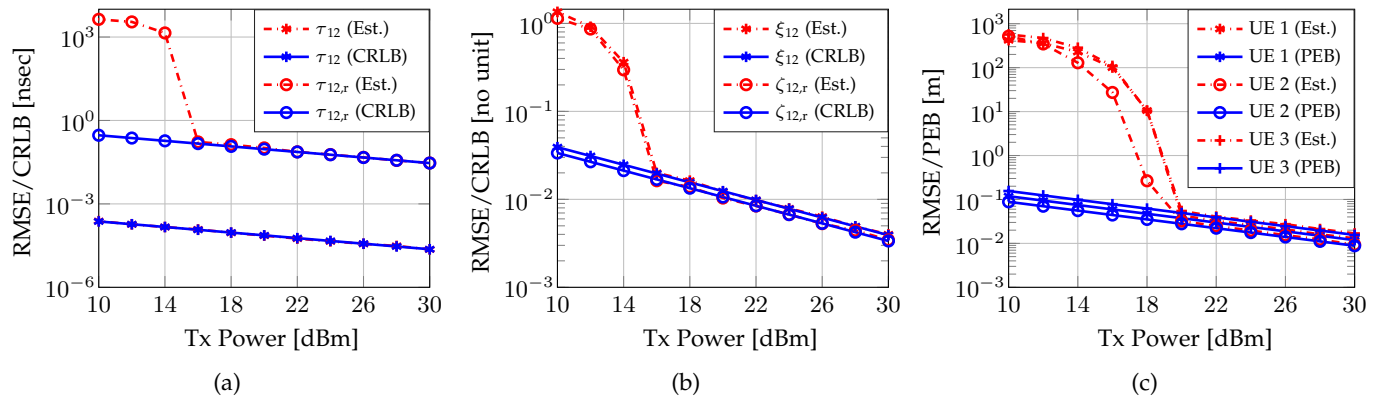
\begin{figure*}[!t]
    \begin{minipage}[b]{0.33\linewidth}
  \centering 
    \centerline{
%
%
\begin{tikzpicture}

\begin{axis}[%
width=44mm,
height=37mm,
at={(0mm,0mm)},
scale only axis,
xmin=10,
xmax=30,
xtick={ 6, 10, 14, 18, 22, 26, 30},
ymode=log,
ymin=1e-06,
ymax=10000,
yminorticks=true,
yticklabel style = {font=\small,xshift=0.5ex},
xticklabel style = {font=\small,yshift=0ex},
axis background/.style={fill=white},
xmajorgrids,
ymajorgrids,
legend style={font=\scriptsize, at={(1.0, 1.0)}, anchor=north east,legend cell align=left, align=left, draw=white!15!black}
]
\addplot [color=red, dashdotted, line width=1.0pt, mark=asterisk, mark options={solid, red}]
  table[row sep=crcr]{%
6	0.000372392338551587\\
8	0.000293562779776075\\
10	0.000240423964667643\\
12	0.000184215151332757\\
14	0.000147437323282433\\
16	0.000120391349414186\\
18	9.17858388877288e-05\\
20	7.23584929600148e-05\\
22	5.66019152570435e-05\\
24	4.50439108662969e-05\\
26	3.67437189976281e-05\\
28	3.08310713316437e-05\\
30	2.3093618353583e-05\\
};
\addlegendentry{$\tau{}_{\text{12}}\text{ (Est.)}$}

\addplot [color=blue, line width=1.0pt, mark=asterisk, mark options={solid, blue}]
  table[row sep=crcr]{%
6	0.000370686530131693\\
8	0.000294446777115579\\
10	0.000233887388686471\\
12	0.000185783356579597\\
14	0.000147572965673023\\
16	0.00011722137331608\\
18	9.31122465381173e-05\\
20	7.39616864238346e-05\\
22	5.87498558142755e-05\\
24	4.66666692592596e-05\\
26	3.70686530131693e-05\\
28	2.94446777115577e-05\\
30	2.33887388686471e-05\\
};
\addlegendentry{$\tau{}_{\text{12}}\text{ (CRLB)}$}

\addplot [color=red, dashdotted, line width=1.0pt, mark=o, mark options={solid, red}]
  table[row sep=crcr]{%
6	4830.05584459132\\
8	4748.0725667999\\
10	4327.91131275779\\
12	3500.27722297902\\
14	1402.03198913736\\
16	0.169031760463819\\
18	0.134835363092136\\
20	0.102957452234524\\
22	0.0742474824778553\\
24	0.0594815174896518\\
26	0.0469727595142329\\
28	0.036575239781067\\
30	0.0294828463749194\\
};
\addlegendentry{$\tau{}_{\text{12,r}}\text{ (Est.)}$}

\addplot [color=blue, line width=1.0pt, mark=o, mark options={solid, blue}]
  table[row sep=crcr]{%
6	0.465417224826361\\
8	0.369694042606596\\
10	0.293658416251781\\
12	0.233261171393206\\
14	0.185285934502483\\
16	0.147177849272596\\
18	0.116907521203218\\
20	0.0928629449433432\\
22	0.0737636591281441\\
24	0.0585925571420623\\
26	0.046541722482636\\
28	0.0369694042606596\\
30	0.0293658416251781\\
};
\addlegendentry{$\tau{}_{\text{12,r}}\text{ (CRLB)}$}

\end{axis}
\node[rotate=0,fill=white] (BOC6) at (2.30cm,-.7cm){\small Tx Power [dBm]};
\node[rotate=90] at (-10mm,18mm){\small RMSE/CRLB [nsec]};
\end{tikzpicture}%
}
    \centerline{\small{(a)}} 
    \end{minipage}
    \begin{minipage}[b]{0.33\linewidth}
  \centering 
    \centerline{
%
%
\begin{tikzpicture}

\begin{axis}[%
width=44mm,
height=37mm,
at={(0mm,0mm)},
scale only axis,
xmin=10,
xmax=30,
xtick={ 6, 10, 14, 18, 22, 26, 30},
ymode=log,
ymin=0.001,
ymax=1.44986827712351,
yminorticks=true,
yticklabel style = {font=\small,xshift=0.5ex},
xticklabel style = {font=\small,yshift=0ex},
axis background/.style={fill=white},
xmajorgrids,
ymajorgrids,
legend style={font=\scriptsize, at={(1.0, 1.0)}, anchor=north east,legend cell align=left, align=left, draw=white!15!black}
]
\addplot [color=red, dashdotted, line width=1.0pt, mark=asterisk, mark options={solid, red}]
  table[row sep=crcr]{%
6	1.44986827712351\\
8	1.42164296356263\\
10	1.35312075855242\\
12	0.916460635733458\\
14	0.362435082200427\\
16	0.0199894309608853\\
18	0.0163169156875251\\
20	0.0122125297386628\\
22	0.00981064624864471\\
24	0.0078943344590048\\
26	0.00635816471711379\\
28	0.00493165165607541\\
30	0.00399641723334212\\
};
\addlegendentry{$\xi{}_{\text{12}}\text{ (Est.)}$}

\addplot [color=blue, line width=1.0pt, mark=asterisk, mark options={solid, blue}]
  table[row sep=crcr]{%
6	0.0621008760382108\\
8	0.0493284792382634\\
10	0.0391830038349632\\
12	0.031124166267421\\
14	0.0247228040484656\\
16	0.019638021297252\\
18	0.015599034790524\\
20	0.0123907537685596\\
22	0.00984232556788318\\
24	0.00781803709391831\\
26	0.00621008760382108\\
28	0.00493284792382635\\
30	0.00391830038349632\\
};
\addlegendentry{$\xi{}_{\text{12}}\text{ (CRLB)}$}

\addplot [color=red, dashdotted, line width=1.0pt, mark=o, mark options={solid, red}]
  table[row sep=crcr]{%
6	1.1242833917467\\
8	1.17254750009782\\
10	1.14175167902874\\
12	0.863738744346097\\
14	0.29674487319448\\
16	0.0162838779630849\\
18	0.0138185822839244\\
20	0.0103509277345815\\
22	0.00840488530160225\\
24	0.00672415801881242\\
26	0.0052965826000459\\
28	0.00443880406608511\\
30	0.00345172439765259\\
};
\addlegendentry{$\zeta{}_{\text{12,r}}\text{ (Est.)}$}

\addplot [color=blue, line width=1.0pt, mark=o, mark options={solid, blue}]
  table[row sep=crcr]{%
6	0.0533082974932606\\
8	0.0423442858439786\\
10	0.0336352618251079\\
12	0.0267174381500269\\
14	0.0212224154820661\\
16	0.0168575638264548\\
18	0.0133904389160198\\
20	0.0106364037063453\\
22	0.00844879577987606\\
24	0.00671111703737491\\
26	0.00533082974932606\\
28	0.00423442858439786\\
30	0.00336352618251079\\
};
\addlegendentry{$\zeta{}_{\text{12,r}}\text{ (CRLB)}$}

\end{axis}
\node[rotate=0,fill=white] (BOC6) at (2.30cm,-.7cm){\small Tx Power [dBm]};
\node[rotate=90] at (-10mm,18mm){\small RMSE/CRLB [no unit]};
\end{tikzpicture}%
}
    \centerline{\small{(b)}} 
    \end{minipage}
    \begin{minipage}[b]{0.33\linewidth}
  \centering 
    \centerline{
%
%
\begin{tikzpicture}

\begin{axis}[%
width=46mm,
height=37mm,
at={(0mm,0mm)},
scale only axis,
xmin=10,
xmax=30,
xtick={ 10, 14, 18, 22, 26, 30},
ytick={0.001, 0.01,0.1,1,10,100,1000},
ymode=log,
ymin=0.001,
yminorticks=true,
yticklabel style = {font=\small,xshift=0.5ex},
xticklabel style = {font=\small,yshift=0ex},
axis background/.style={fill=white},
xmajorgrids,
ymajorgrids,
legend style={font=\scriptsize, at={(1.0, 1.0)}, anchor=north east,legend cell align=left, align=left, draw=white!15!black}
]
\addplot [color=red, dashdotted, line width=1.0pt, mark=asterisk, mark options={solid, red}]
  table[row sep=crcr]{%
6	433.381455147191\\
8	425.091684672797\\
10	421.210107617702\\
12	359.805761617397\\
14	227.717850058202\\
16	100.388553828376\\
18	10.5544043000861\\
20	0.0430976726627546\\
22	0.0325904106708247\\
24	0.026098213897482\\
26	0.0208893528145579\\
28	0.016027699940044\\
30	0.0130516240416269\\
};
\addlegendentry{UE 1 (Est.)}

\addplot [color=blue, line width=1.0pt, mark=asterisk, mark options={solid, blue}]
  table[row sep=crcr]{%
6	0.187837011594612\\
8	0.14920424185166\\
10	0.118517142059126\\
12	0.0941415122132441\\
14	0.0747792612228019\\
16	0.059399278558407\\
18	0.047182524086516\\
20	0.0374784110652269\\
22	0.0297701601033512\\
24	0.0236472787191111\\
26	0.0187837011605146\\
28	0.014920424185447\\
30	0.0118517142053024\\
};
\addlegendentry{UE 1 (PEB)}

\addplot [color=red, dashdotted, line width=1.0pt, mark=o, mark options={solid, red}]
  table[row sep=crcr]{%
6	590.836827471127\\
8	583.845768708457\\
10	518.881167976177\\
12	346.221984124834\\
14	129.155763729395\\
16	27.2303816618286\\
18	0.266243538744891\\
20	0.0321298926025323\\
22	0.0249099574416191\\
24	0.0196402519217112\\
26	0.0154014690667197\\
28	0.0122701734071545\\
30	0.00934526959892422\\
};
\addlegendentry{UE 2 (Est.)}

\addplot [color=blue, line width=1.0pt, mark=o, mark options={solid, blue}]
  table[row sep=crcr]{%
6	0.139934183749028\\
8	0.111153673175372\\
10	0.0882925010116706\\
12	0.0701332264349017\\
14	0.0557088019647708\\
16	0.0442510743229577\\
18	0.0351498777558631\\
20	0.0279205403470077\\
22	0.0221780735280433\\
24	0.0176166699904073\\
26	0.0139934183763313\\
28	0.0111153673175613\\
30	0.00882925009927041\\
};
\addlegendentry{UE 2 (PEB)}

\addplot [color=red, dashdotted, line width=1.0pt, mark=+, mark options={solid, red}]
  table[row sep=crcr]{%
6	583.224578451355\\
8	575.278799625086\\
10	567.866477622\\
12	471.463494779319\\
14	284.098494187577\\
16	109.513190739693\\
18	10.4030265056126\\
20	0.0546143421751581\\
22	0.0406857500508233\\
24	0.033112070701675\\
26	0.0273610902957576\\
28	0.0212011809465362\\
30	0.0168223852643038\\
};
\addlegendentry{UE 3 (Est.)}

\addplot [color=blue, line width=1.0pt, mark=+, mark options={solid, blue}]
  table[row sep=crcr]{%
6	0.24671841240152\\
8	0.195975401014473\\
10	0.1556687943521\\
12	0.123652118598691\\
14	0.0982203690994532\\
16	0.0780192123963258\\
18	0.0619728632630764\\
20	0.0492267950738362\\
22	0.0391022332335536\\
24	0.031060007895379\\
26	0.0246718412415926\\
28	0.0195975401022356\\
30	0.015566879435216\\
};
\addlegendentry{UE 3 (PEB)}

\end{axis}
\node[rotate=0,fill=white] (BOC6) at (2.35cm,-.7cm){\small Tx Power [dBm]};
\node[rotate=90] at (-10mm,18mm){\small RMSE/PEB [m]};
\end{tikzpicture}%
}
    \centerline{\small{(c)}} 
    \end{minipage}
    \caption{Comparison of the RMSE and CRLB/PEB for (a) time delays, (b) spatial frequencies and (c) position estimates, as a function of the transmission power.}
    \label{Fig:estimator_power} \vspace{-5mm}
\end{figure*}

For ease of illustration, we only focus on the case where UE $1$ is the transmitter and UE $2$ is the receiver to evaluate the delay and spatial frequency estimators since there are many channel parameters involved and the results are similar. Figure~\ref{Fig:estimator_power}(a) shows the performance of the time delay estimator for the LoS path and the RIS path (i.e., $\tau_{12}$ and $\tau_{12,r}$) in comparison to the CRLB. As seen in the figure, the LoS delay estimation error attains the CRLB for all transmission power values due to the large channel gain value as observed in \eqref{gain_1}. However, in the case of the RIS path delay, the channel gain value \eqref{gain_2} is much lower and thus there is a high estimation error when the power is below $16$ dBm since the RIS signal has a low signal-to-noise ratio (SNR). The RMSE of the RIS path delay gets very close to the CRLB when the transmission power is $16$ dBm or higher. Figure~\ref{Fig:estimator_power}(b) compares the estimation RMSE values of the spatial frequencies $\xi_{12}$ and $\zeta_{12}$ with the CRLB. Since the spatial frequencies are only included in the RIS path, we see similar performance to the RIS path delay where the error is relatively large if the power is below $16$ dBm and small otherwise.  

Figure~\ref{Fig:estimator_power}(c) shows the RMSE of the positioning algorithm for all three UEs in comparison to the PEBs. Again, due to the large estimation error of the channel parameters when the power is below $16$ dBm, the positioning errors are also large. The positioning errors are in the range of hundreds of meters with low transmission power since we do not account for the geometry of the environment in our proposed localization algorithm. On the other hand, the positioning error gets close to the PEB for all the UEs when the transmission power is $18$ dBm or higher. We can see from Figure~\ref{Fig:estimator_power}(c) that the proposed algorithm can provide centimeter-level accuracy at $20$ dBm transmission power.

\subsection{Replacing RIS with AP}
In this section, we study the effect of replacing the RIS (i.e., passive element) with an AP (i.e., active source) on the positioning performance. For the RIS scenario, we consider the same parameters described in Section~\ref{sec:simulation_setup}. As a comparison scenario, we replace the RIS with an AP at the same location. The AP parameters are: $10$ dBm transmission power, $3\times3$ array and $12$ transmissions. In both scenarios, we fix the total number of transmissions to 120. The number of transmission per UE in the RIS scenario is 40 while it is 36 per UE in the AP scenario. In the AP case, we assume that the UEs communicate with each other directly using the same parameters as the RIS scenario. In the AP scenario, each UE estimates the LoS time delays from its neighboring UEs. In addition, the AP estimates the time delays from each of the UE in addition to the angle-of-arrivals (AoAs). We use the random phase codebook for the RIS profile design. 

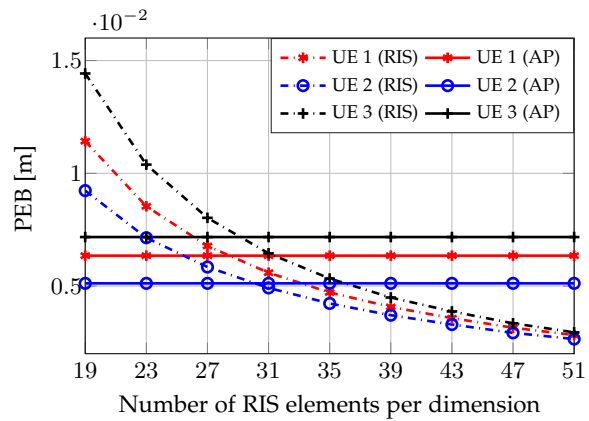
\begin{figure}[!t]
  \centering  
\centerline{
%
%
\begin{tikzpicture}

\begin{axis}[%
width=65mm,
height=42mm,
at={(0mm,0mm)},
scale only axis,
xmin=19,
xmax=51,
xtick={19, 23, 27, 31, 35, 39, 43, 47, 51},
yticklabel style = {font=\small,xshift=0.5ex},
xticklabel style = {font=\small,yshift=0ex},
ymin=0.002,
ymax=0.016,
axis background/.style={fill=white},
xmajorgrids,
ymajorgrids,
legend columns=2, 
legend style={font=\scriptsize, at={(1.0, 1.0)}, anchor=north east, legend cell align=left, align=left, draw=white!15!black}
]
\addplot [color=red, dashdotted, line width=1.0pt, mark=asterisk, mark options={solid, red}]
  table[row sep=crcr]{%
19	0.0114248112581046\\
23	0.00853702938978799\\
27	0.0067782036719681\\
31	0.00559249691081777\\
35	0.00471814864306255\\
39	0.00407466371381864\\
43	0.00357801620999462\\
47	0.00315372264045803\\
51	0.00282817148419735\\
};
\addlegendentry{UE 1 (RIS)}

\addplot [color=red, mark=asterisk, mark options={solid, red}, line width=1.0pt]
  table[row sep=crcr]{%
19	0.00634861753157681\\
23	0.00634861753157681\\
27	0.00634861753157681\\
31	0.00634861753157681\\
35	0.00634861753157681\\
39	0.00634861753157681\\
43	0.00634861753157681\\
47	0.00634861753157681\\
51	0.00634861753157681\\
};
\addlegendentry{UE 1 (AP)}

\addplot [color=blue, dashdotted, line width=1.0pt, mark=o, mark options={solid, blue}]
  table[row sep=crcr]{%
19	0.00923388129801377\\
23	0.00714264053196883\\
27	0.00584535700293816\\
31	0.00491797359448578\\
35	0.00422889392396607\\
39	0.00370838461461778\\
43	0.00329349991278927\\
47	0.00292639465516553\\
51	0.00265084048236345\\
};
\addlegendentry{UE 2 (RIS)}

\addplot [color=blue, line width=1.0pt, mark=o, mark options={solid, blue}]
  table[row sep=crcr]{%
19	0.00512013666957502\\
23	0.00512013666957502\\
27	0.00512013666957502\\
31	0.00512013666957502\\
35	0.00512013666957502\\
39	0.00512013666957502\\
43	0.00512013666957502\\
47	0.00512013666957502\\
51	0.00512013666957502\\
};
\addlegendentry{UE 2 (AP)}

\addplot [color=black, dashdotted, line width=1.0pt, mark=+, mark options={solid, black}]
  table[row sep=crcr]{%
19	0.0144273161339798\\
23	0.0103865924786006\\
27	0.0080256052930701\\
31	0.00645131677330265\\
35	0.00533366260977136\\
39	0.00449276188534623\\
43	0.00387389818059407\\
47	0.00334864104597745\\
51	0.00294600805405988\\
};
\addlegendentry{UE 3 (RIS)}

\addplot [color=black, line width=1.0pt, mark=+, mark options={solid, black}]
  table[row sep=crcr]{%
19	0.00717333793770305\\
23	0.00717333793770305\\
27	0.00717333793770305\\
31	0.00717333793770305\\
35	0.00717333793770305\\
39	0.00717333793770305\\
43	0.00717333793770305\\
47	0.00717333793770305\\
51	0.00717333793770305\\
};
\addlegendentry{UE 3 (AP)}

\end{axis}

\node[rotate=0,fill=white] (BOC6) at (3.25cm,-.7cm){\small Number of RIS elements per dimension};
\node[rotate=90] at (-8mm,22mm){\small PEB [m]};
\end{tikzpicture}%
}
\caption{PEBs of RIS-enabled versus AP-aided CP for different RIS sizes.}
\label{Fig:ris_ap_bound} \vspace{-5mm}
\end{figure}

Figure~\ref{Fig:ris_ap_bound} shows the PEBs for the three UEs under the RIS-enabled and AP-aided CP as a function of the number of RIS elements per dimension increases. Here, we vary the RIS size from $19\times19$ elements to $51\times51$ elements. For accurate representation, each value is averaged over $100$ Monte-Carlo simulations. We can see that the PEBs of the AP scenario are constant lines since there is no RIS involved. It is also clear that the PEBs of the RIS-enabled scenario decrease as the RIS size increases. With a sufficiently large RIS ($35\times 35$ array), we can replace the AP with a RIS without performance degradation. The far-field conditions are satisfied when the distance between the RIS center and the closest UE is much greater than the maximum distance between the RIS center and one RIS element. In other words,  
\begin{equation}
    \min_{i} {\left \lVert \pv_\text{R} - \pv_{\text{U}_i}  \right \rVert}_2 \gg \max_m {\left \lVert \pv_\text{R} - \pv_{\text{R},m}  \right \rVert}_2. 
    \label{eq:far-field}
\end{equation}
When the RIS size is $35\times35$, the far-field conditions are satisfied. When the RIS size is very large (e.g., $500\times500$), \eqref{eq:far-field} is not satisfied anymore and have to consider near-field localization which is beyond the scope of this paper.

\subsection{Effect of RIS Profile and Power Allocation on Positioning Performance}
In this section, we evaluate the positioning performance under different RIS profiles and different power allocation. We consider two cases for the phase shifts at the RIS: random phase codebook and directional codebook. 

\begin{figure}[!t]
  \centering  
\centerline{\input{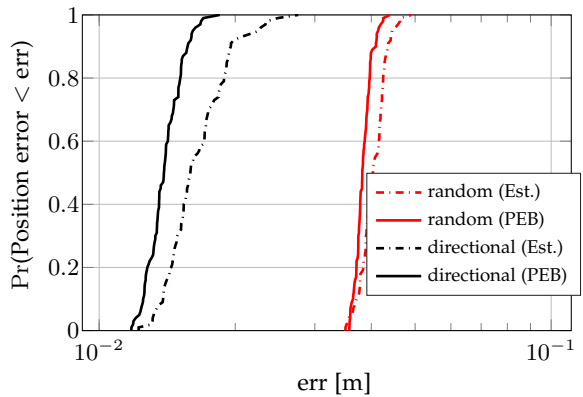}}
\caption{ECDF of the positioning error of UE $3$ for the lower bound and the estimator for two RIS profiles.}
\label{Fig:ris_profile_cdf} \vspace{-5mm}
\end{figure}

Figure~\ref{Fig:ris_profile_cdf} shows the empirical cumulative distribution function (ECDF) for both the estimator and the PEB for two RIS profiles. For ease of illustration, we only show the performance of UE $3$. Here, we set the number of RIS elements to $M=121$ and the total transmission power is evenly distributed between the UEs. Prior information about the UEs' positions is needed in the case of the directional codebook. We assume that the distribution of the prior information of UE $i$ is normal with $\mathcal{N}(\pv_{\text{U}_i}, \Sigmam_{\pv_{\text{U}_i}})$. We set $\Sigmam_{\pv_{\text{U}_i}}=1.5\mathbf{I}_3$ in Figure~\ref{Fig:ris_profile_cdf}. To plot the ECDF, we generate $100$ different codebooks in each case. To accurately evaluate the estimator performance, we take the average of $200$ noise realizations for each codebook. It is clear that the directional codebook outperforms the random phase codebook since the energy is directed toward the UEs' positions. 

\begin{figure}[!t]
\begin{minipage}[b]{0.9\linewidth}
  \centering
\centerline{\includegraphics[width=0.85\linewidth]{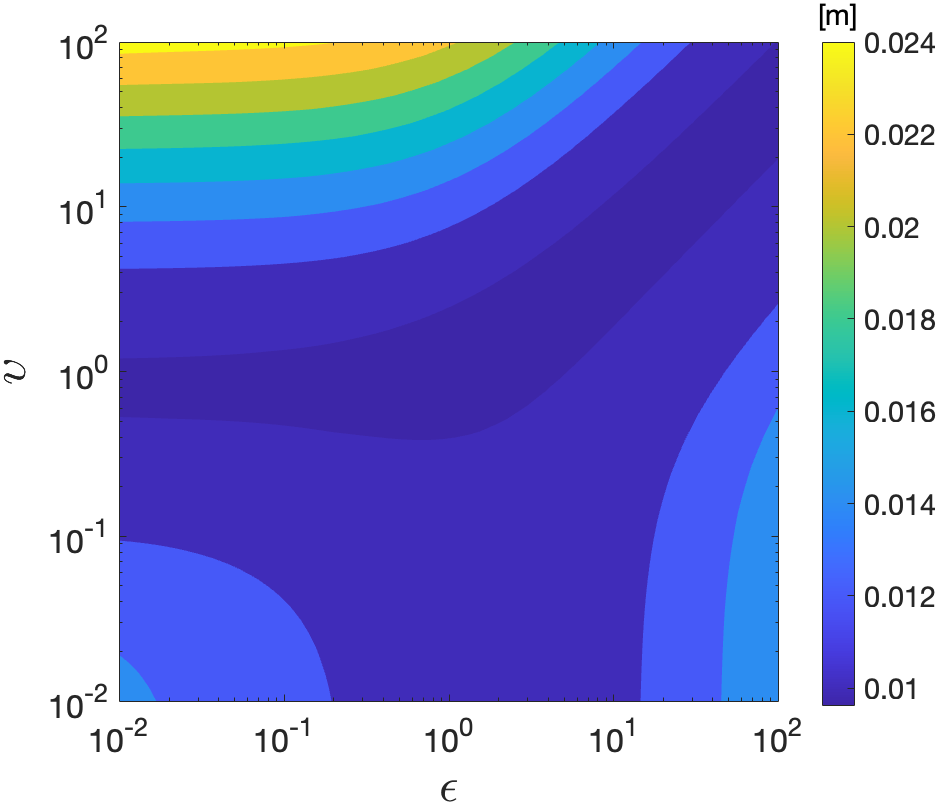}}
\centerline{\small{(a)}} \medskip
\end{minipage}
\begin{minipage}[b]{0.9\linewidth}
  \centering
\centerline{\includegraphics[width=0.85\linewidth]{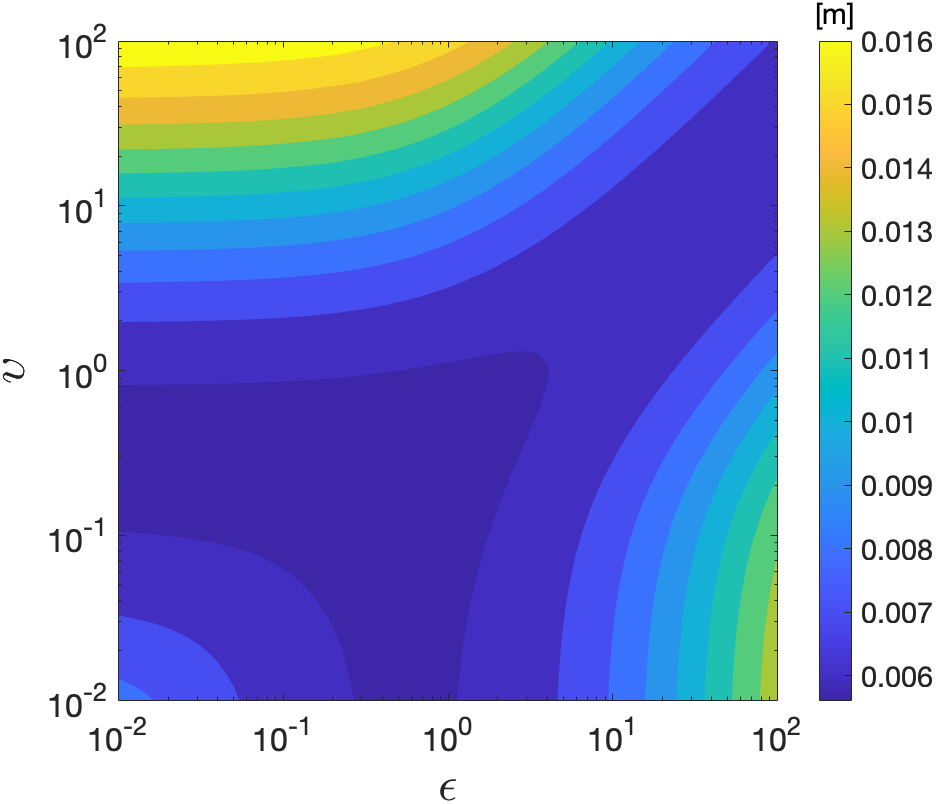}}
\centerline{\small{(b)}} \medskip
\end{minipage}
\caption{The average PEB of the three UEs as a function of the power scaling factors $\epsilon$ and $\upsilon$ for a fixed RIS profile for two cases of UEs' positions. In both cases, $\pv_{\text{UE}_1}=[4, 3, -1]^\top$ m and $\pv_{\text{UE}_2}=[4.5, 1, -0.5]^\top$ m. In (a), $\pv_{\text{UE}_3}=[5, -3, -1]^\top$ m. In (b), $\pv_{\text{UE}_3}=[1, -1, -1]^\top$ m.  }
\label{Fig:heat_power} \vspace{-5mm}
\end{figure}

To evaluate the power allocation strategy on the positioning performance, we define $P_2 = \epsilon P_1$ and $P_3 = \upsilon P_1$ where $\epsilon,~\upsilon >0$ are scaling factors. Then, $P_1+P_2+P_3=(1+\epsilon+\upsilon)P_1=P_{\text{tot}}$ and thus $P_1 = 1/(1+\epsilon+\upsilon)P_{\text{tot}}$. Hence, the transmission power of each UE can be defined in terms of $\epsilon$, $\upsilon$ and $P_{\text{tot}}$. We observe that $P_{\text{tot}}$ is evenly distributed between the UEs when $\epsilon=1$ and $\upsilon=1$. Here, we consider two cases for the UEs' positions. In the first case, we use the positions defined in Section~\ref{sec:simulation_setup}. In the second case, we move UE $3$ from $\pv_{\text{UE}_3}=[5, -3, -1]^\top$ m to $\pv_{\text{UE}_3}=[1, -1, -1]^\top$ m (i.e., closer to the RIS) while we keep the positions of UEs $1$ and $2$ the same as the first case. Figures~\ref{Fig:heat_power}(a) and \ref{Fig:heat_power}(b) show the average PEB of the three UEs (i.e., $1/3\sum_{i=1}^3 \text{PEB}_i$) as a function of the power scaling factors $\epsilon$ and $\upsilon$. Here, we set $P_{\text{tot}}=600$ mW and we use one realization of the  directional profile with $\Sigmam_{\pv_{\text{U}_i}}=0.5\mathbf{I}_3$. As seen in both figures, it is clear that there are three corners with high average PEB: bottom left (power allocated to UE $1$ only), bottom right (power allocated to UE $2$ only) and top left (power allocated to UE $3$ only). In Figure~\ref{Fig:heat_power}(a), we observe a small area centered around $\epsilon=1$ and $\upsilon=1$ where the PEB value is small which suggests that a uniform power allocation between the UEs is near optimal in the first case. However, in Figure~\ref{Fig:heat_power}(b), the values of both $\epsilon$ and $\upsilon$ are less than $1$ in the optimal PEB area. This suggests that uniform power allocation is not optimal in the the second case where we observe that UE $1$ is allocated more power since it is the farthest from the RIS.

\begin{figure}[!t]
  \centering  
\centerline{
%
%
\definecolor{mycolor1}{rgb}{1.00000,0.00000,1.00000}%
\begin{tikzpicture}

\begin{axis}[%
width=65mm,
height=42mm,
at={(0mm,0mm)},
scale only axis,
xmode=log,
xmin=0.001,
xmax=100,
xminorticks=true,
ymode=log,
ymin=0.0075,
ymax=0.1,
yminorticks=true,
xmajorgrids,
ymajorgrids,
legend style={font=\scriptsize, at={(0.83, 1.1)}, anchor=north east, legend cell align=left, align=left, draw=white!15!black}
]
\addplot [color=red, line width=1.0pt, mark=asterisk, mark options={solid, red}]
  table[row sep=crcr]{%
0.001	0.0300216988370489\\
0.00242446201708233	0.0300216988370489\\
0.00587801607227491	0.0300216988370489\\
0.01425102670303	0.0300216988370489\\
0.0345510729459222	0.0300216988370489\\
0.0837677640068292	0.0300216988370489\\
0.203091762090473	0.0300216988370489\\
0.492388263170674	0.0300216988370489\\
1.19377664171444	0.0300216988370489\\
2.89426612471675	0.0300216988370489\\
7.01703828670383	0.0300216988370489\\
17.0125427985259	0.0300216988370489\\
41.2462638290135	0.0300216988370489\\
100	0.0300216988370489\\
};
\addlegendentry{random, $\epsilon=\upsilon=1$}

\addplot [color=blue, line width=1.0pt, mark=o, mark options={solid, blue}]
  table[row sep=crcr]{%
0.001	0.0571398891350399\\
0.00242446201708233	0.0423226451835136\\
0.00587801607227491	0.0298865315621201\\
0.01425102670303	0.0206833415894586\\
0.0345510729459222	0.0145910421837544\\
0.0837677640068292	0.00997639875972858\\
0.203091762090473	0.00858039440037684\\
0.492388263170674	0.00849217152210724\\
1.19377664171444	0.0101256185825874\\
2.89426612471675	0.0135100860660264\\
7.01703828670383	0.0168540090666618\\
17.0125427985259	0.0214484082867963\\
41.2462638290135	0.0253290451864307\\
100	0.0265619277303669\\
};
\addlegendentry{directional, $\epsilon=\upsilon=1$}

\addplot [color=black, line width=1.0pt, mark=+, mark options={solid, black}]
  table[row sep=crcr]{%
0.001	0.0562591002063152\\
0.00242446201708233	0.0475337972284188\\
0.00587801607227491	0.0401562335955538\\
0.01425102670303	0.032297688599063\\
0.0345510729459222	0.0235062742567276\\
0.0837677640068292	0.0172006109806624\\
0.203091762090473	0.0151614329298525\\
0.492388263170674	0.0149148793719906\\
1.19377664171444	0.0174321301599113\\
2.89426612471675	0.0230277099744093\\
7.01703828670383	0.0289485437918566\\
17.0125427985259	0.0398882215963525\\
41.2462638290135	0.0473691388712325\\
100	0.0491568474434\\
};
\addlegendentry{directional, $\epsilon=1$, $\upsilon=40$}

\addplot [color=cyan, line width=1.0pt, mark=diamond, mark options={solid, cyan}]
  table[row sep=crcr]{%
0.001	0.0648865586170148\\
0.00242446201708233	0.0475285585558812\\
0.00587801607227491	0.0344674489127432\\
0.01425102670303	0.0235678677638717\\
0.0345510729459222	0.0166542051756696\\
0.0837677640068292	0.0118060052285726\\
0.203091762090473	0.0110255275981428\\
0.492388263170674	0.0114941218697389\\
1.19377664171444	0.0148164798748186\\
2.89426612471675	0.0198873915016921\\
7.01703828670383	0.0273781772537898\\
17.0125427985259	0.0305562339851937\\
41.2462638290135	0.0353146677121355\\
100	0.039840781496142\\
};
\addlegendentry{directional, $\epsilon=50$, $\upsilon=0.4$}

\addplot [color=mycolor1, line width=1.0pt, mark=square, mark options={solid, mycolor1}]
  table[row sep=crcr]{%
0.001	0.0519999695952946\\
0.00242446201708233	0.0390047677350058\\
0.00587801607227491	0.0269857230347884\\
0.01425102670303	0.0187499947839652\\
0.0345510729459222	0.0133542156475093\\
0.0837677640068292	0.00942117239462246\\
0.203091762090473	0.00812694551040243\\
0.492388263170674	0.00817048693905597\\
1.19377664171444	0.00974925765651505\\
2.89426612471675	0.0127104906539708\\
7.01703828670383	0.0159855591532851\\
17.0125427985259	0.0201660528762814\\
41.2462638290135	0.0240609162124213\\
100	0.0254771966389677\\
};
\addlegendentry{directional, optimal power}

\end{axis}

\node[rotate=0,fill=white] (BOC6) at (3.35cm,-.7cm){\small $\sigma_{\text{unc}}^2$ [$\text{m}^2$]};
\node[rotate=90] at (-8mm,22mm){\small PEB [m]};
\end{tikzpicture}%
}
\caption{The average PEB of the three UEs for random and directional profiles as a function of the uncertainty of the prior information with different power allocation strategies.}
\label{Fig:ris_profile_unceratinty} \vspace{-5mm}
\end{figure}
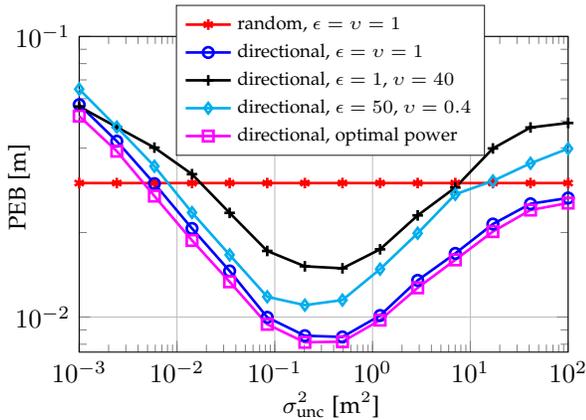

Next, we study how the uncertainty of the prior information of the UEs' positions affect the average PEB. To study the effect of the uncertainty, we define the covariance matrix of the distribution of the prior information as $\Sigmam_{\pv_{\text{U}_i}}=\sigma_{\text{unc}}^2\mathbf{I}_3$ and we vary $\sigma_{\text{unc}}^2$ from $10^{-3}$ to $100$ $\text{m}^2$. Here, there are $M=121$ reflecting elements at the RIS and we use the UEs' positions defined in Section~\ref{sec:simulation_setup}. Figure~\ref{Fig:ris_profile_unceratinty} shows the the average PEB for random and directional RIS profiles as the uncertainty of the prior information increases. Again, the results are averaged over $100$ different RIS profiles. We use uniform power allocation for the random codebook while we adopt different power allocation strategies for the directional codebook. We observe that the random codebook is a constant line since it does not depend on the prior information. We can see from the figure that the directional profile performance is relatively poor when the uncertainty value is large. The positioning performance of the directional profile improves when the uncertainty level is low. However, when the uncertainty value is too small, the performance of the directional profile is worse than the random profile. This happens because when the prior error is too small the RIS reflects the beams to small areas and hence there is less spatial diversity. This could be mitigated by finding an optimal RIS profile by jointly optimizing \eqref{eq_power_allocation_optimization} over transmission powers and RIS control matrices. However, this joint optimization problem is non-convex and will be left for future work. As seen in Figure~\ref{Fig:ris_profile_unceratinty}, it is clear that the positioning performance is affected by the power allocation strategy where the optimal power allocation strategy in \eqref{eq_power_allocation_optimization} leads to the best positioning performance followed by uniform power allocation with a very small margin.

\subsection{Effect of RIS Position on Positioning Performance}
\begin{figure}[!t]
\centering
\includegraphics[width=0.38\textwidth]{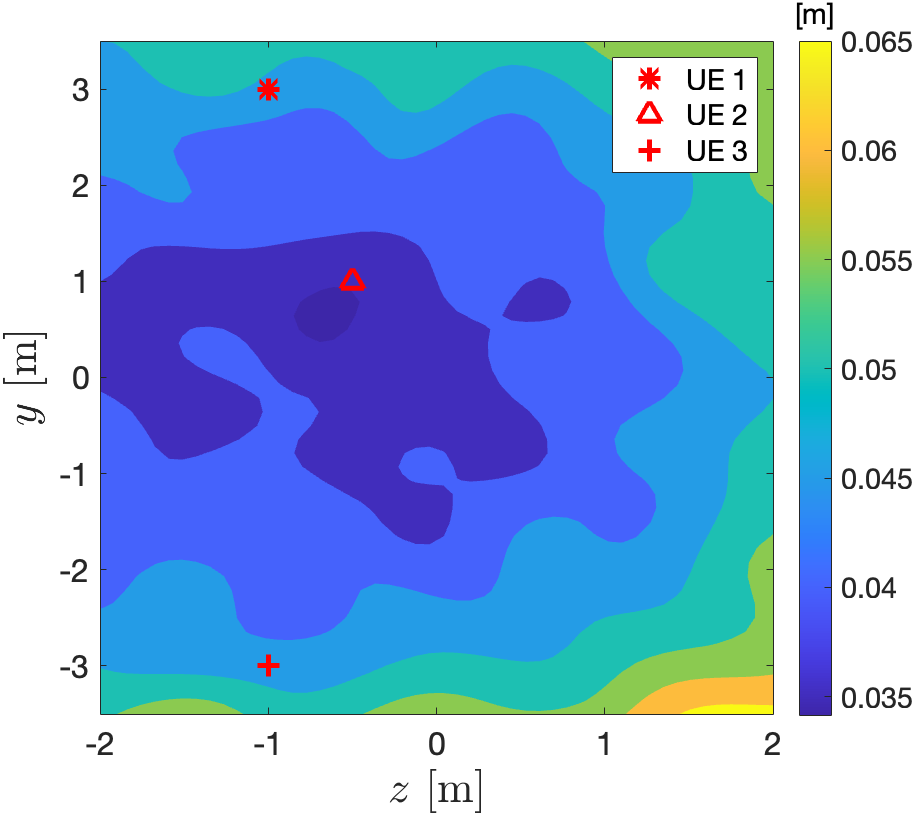}
\caption{The PEB of UE $3$ as the RIS location varies while the UEs' positions are fixed. }
\label{Fig:heat_ris_move} \vspace{-5mm}
\end{figure}

To study the effect of the RIS position on the positioning performance, we fix the locations of the three UEs at $\pv_{\text{UE}_1}=[4, 3, -1]^\top$ m, $\pv_{\text{UE}_2}=[4.5, 1, -0.5]^\top$ m and $\pv_{\text{UE}_3}=[5, -3, -1]^\top$ m. We place the RIS along the $yz$-plane at $\pv_{\text{R}}=[0,y,z]^\top$ m and we vary its $y$ and $z$ values. Figure~\ref{Fig:heat_ris_move} shows the PEB of UE $3$ as the location of the RIS varies on the plane. We can see that the optimal PEB area is centered at $y\approx 0.5$ m and $z\approx -0.5$ m. This center is approximately the average of UEs' positions on the $y$-axis and $z$-axis. The performance degrades when the $|z|$ value is high (i.e., high or low elevation) or when $|y|$ is relatively large.

\begin{figure*}[!t]
    \begin{minipage}[b]{0.33\linewidth}
  \centering 
    \includegraphics[width=\textwidth]{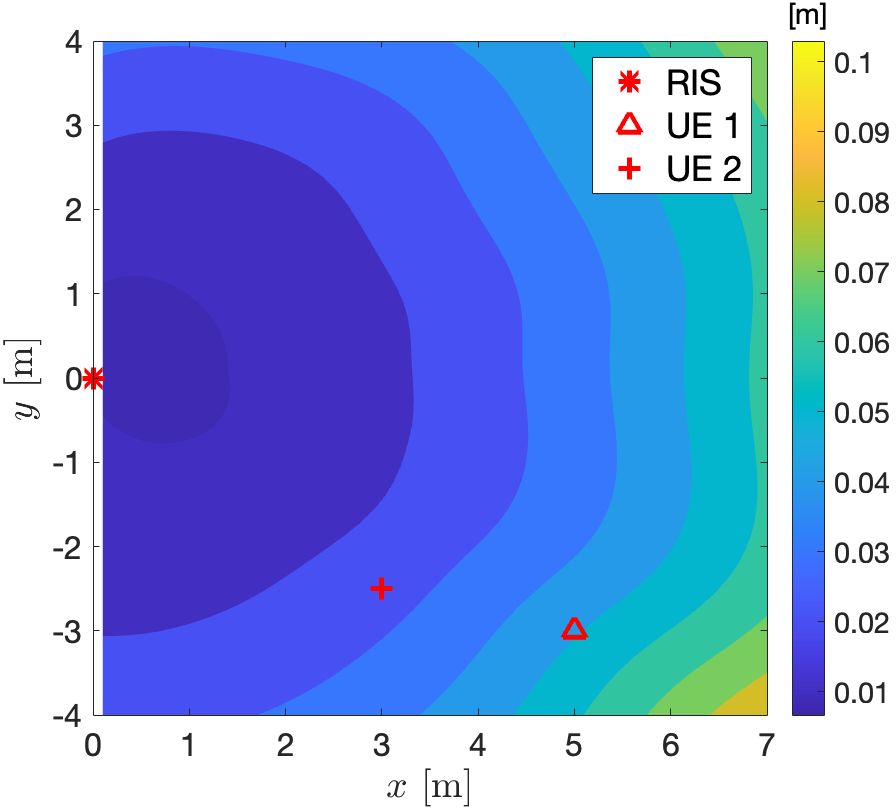}
    \centerline{\small{(a) Random}} 
    \vfill
    \includegraphics[width=\textwidth]{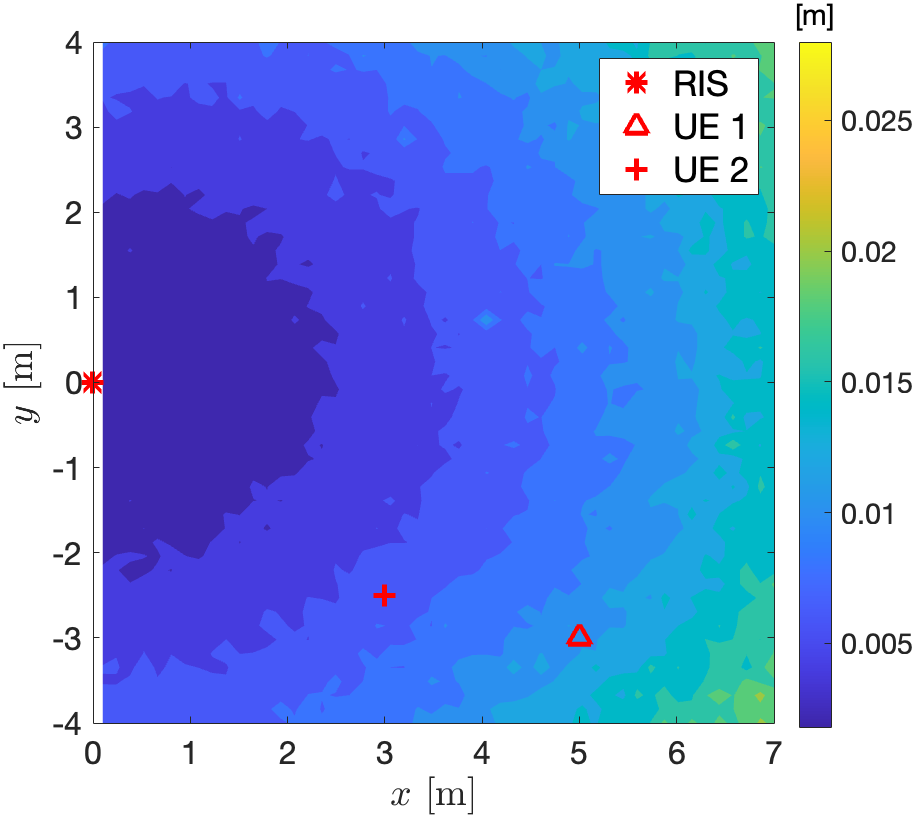}
    \centerline{\small{(b) Directional}}
    \end{minipage}
    \begin{minipage}[b]{0.33\linewidth}
  \centering 
   \includegraphics[width=\textwidth]{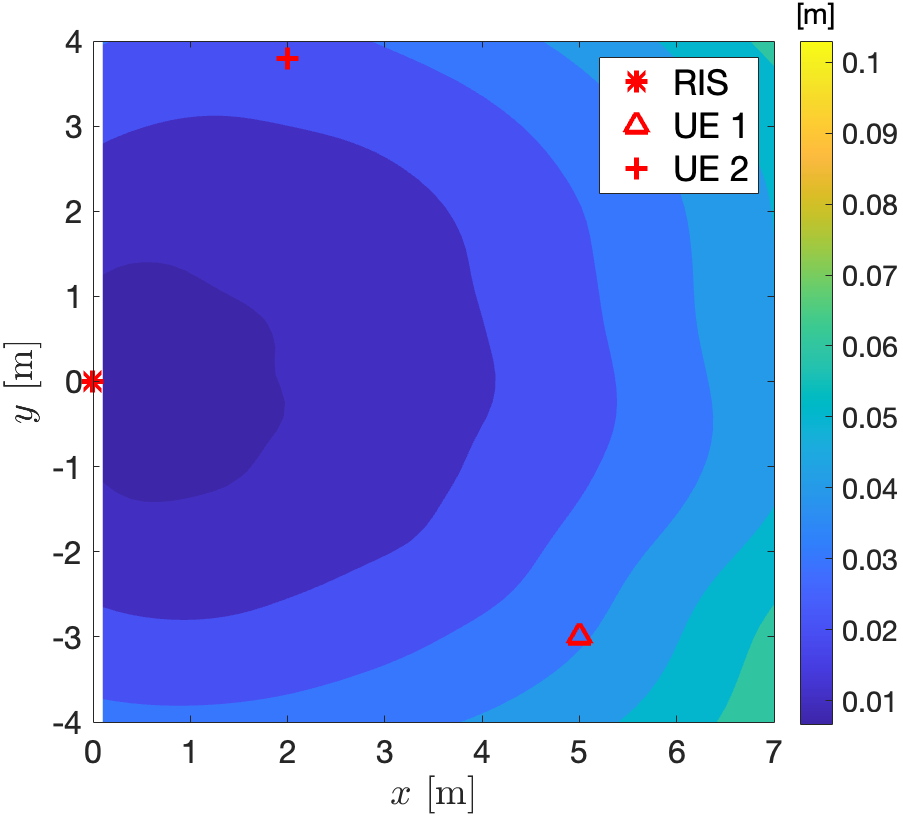}
    \centerline{\small{(c) Random}} 
    \vfill
    \includegraphics[width=\textwidth]{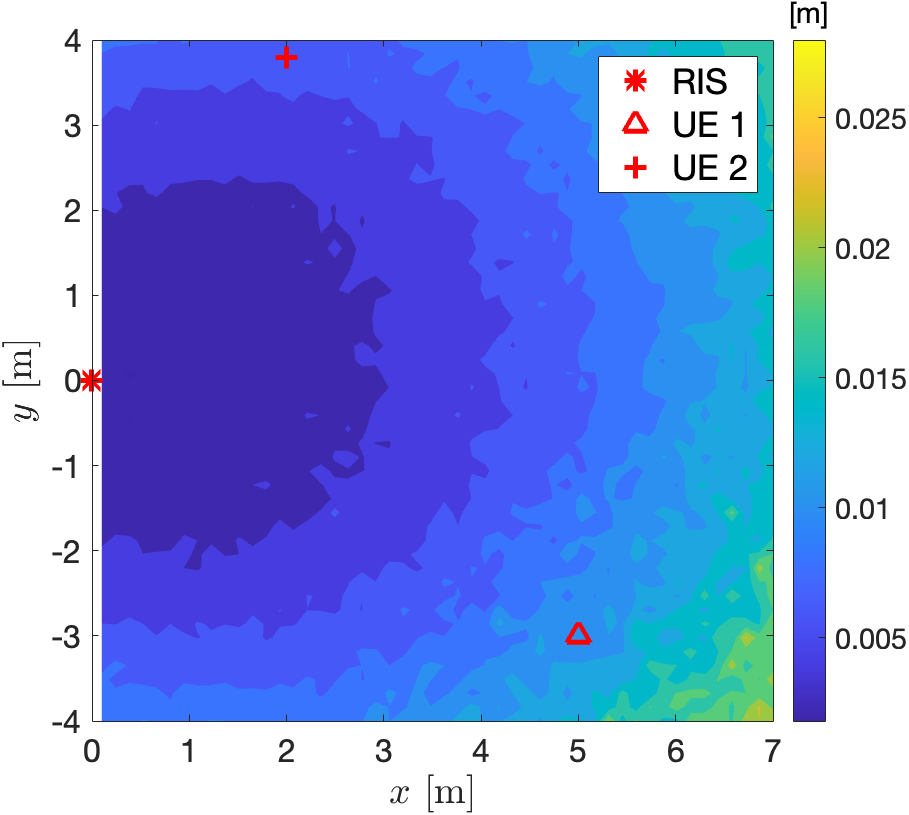}
    \centerline{\small{(d) Directional}}
    \end{minipage}
    \begin{minipage}[b]{0.33\linewidth}
  \centering 
   \includegraphics[width=\textwidth]{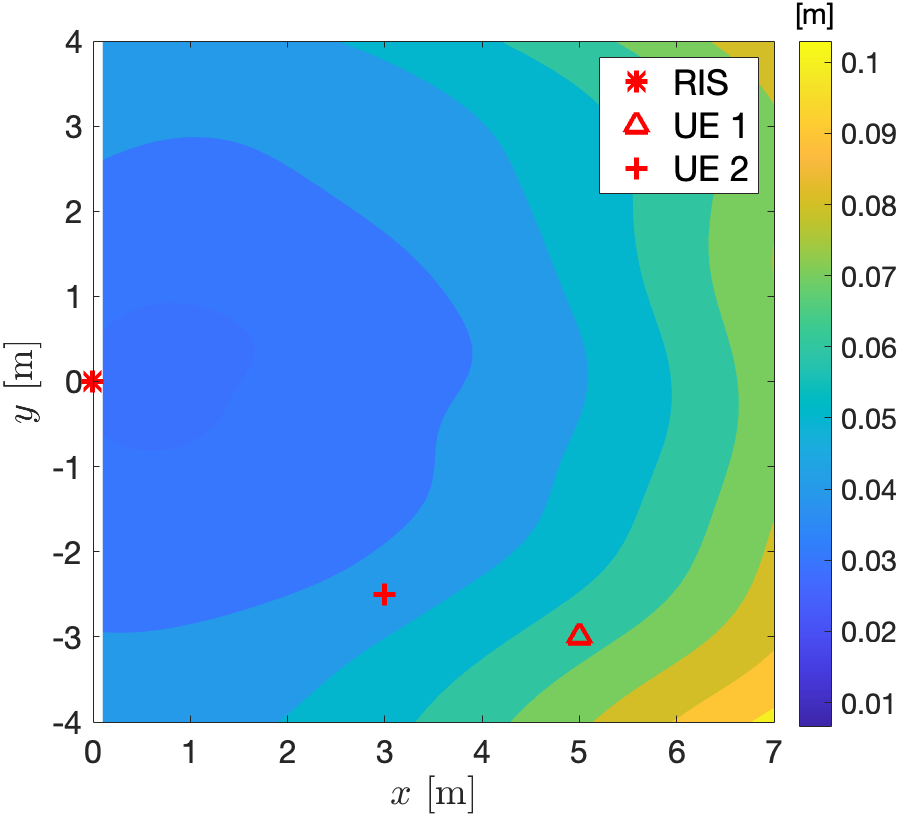}
    \centerline{\small{(e) Random}} 
    \vfill
    \includegraphics[width=\textwidth]{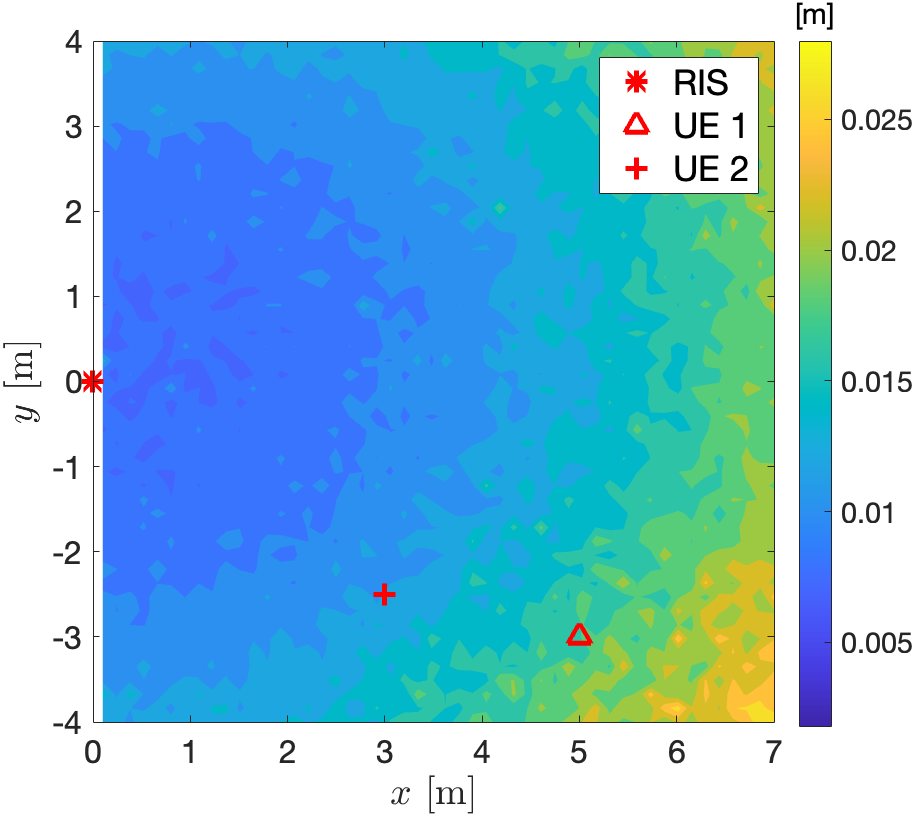}
    \centerline{\small{(f) Directional}}
    \end{minipage}
    \caption{The PEB of UE $3$ under different placements of UEs $1$ and $2$ for both random and directional profiles with the RIS located at $\pv_{\text{R}}=[0,0,0]^\top$ m. In (a) and (b),  $\pv_{\text{UE}_1}=[5, -3, -1.5]^\top$ m, $\pv_{\text{UE}_2}=[3, -2.5, -1]^\top$ m and $\pv_{\text{UE}_3}=[x, y, 1]^\top$ m. In (c) and (d), $\pv_{\text{UE}_1}=[5, -3, -1.5]^\top$ m, $\pv_{\text{UE}_2}=[2, 3.8, -1]^\top$ m and $\pv_{\text{UE}_3}=[x, y, 1]^\top$ m. In (e) and (f), $\pv_{\text{UE}_1}=[5, -3, -1.5]^\top$ m, $\pv_{\text{UE}_2}=[3, -2.5, -1]^\top$ m and $\pv_{\text{UE}_3}=[x, y, 4]^\top$ m.   }
    \label{fig:heat_move_UE3} \vspace{-5mm}
\end{figure*}

\subsection{Effect of UEs' Positions on Positioning Performance}
To study the effect of the UEs' positions on the positioning performance, we fix the RIS location at $\pv_{\text{R}}=[0,0,0]^\top$ m. We also fix the locations of UEs $1$ and $2$ while we vary the location of UE $3$ and calculate its PEB. We consider three scenarios in this section where we vary the location of UE $3$ in the intervals $[0.1,7]$ m and $[-4,4]$ m along the $x$-axis and $y$-axis, respectively. In the first scenario, we place UE $1$, UE $2$ and UE $3$ at $\pv_{\text{UE}_1}=[5, -3, -1.5]^\top$ m, $\pv_{\text{UE}_2}=[3, -2.5, -1]^\top$ m and $\pv_{\text{UE}_3}=[x, y, 1]^\top$ m, respectively. The first scenario is used as a benchmark in this section. In the second scenario, we move UE $2$ to $\pv_{\text{UE}_2}=[2, 3.8, -1]^\top$ (i.e., further away from UE $1$) and we keep the same parameters for UEs $1$ and $3$ as in the first scenario. In the third scenario, we increase the height of UE $3$ and place at $\pv_{\text{UE}_3}=[x, y, 4]^\top$ m, while we keep the locations of UEs $1$ and $2$ the same as the first scenario. For each scenario, we consider two RIS profiles: random phase codebook and directional codebook. Finally, for the directional profile, we use $\Sigmam_{\pv_{\text{U}_i}}=0.5\mathbf{I}_3$. 

Figures~\ref{fig:heat_move_UE3}(a) and \ref{fig:heat_move_UE3}(b) show the PEB for UE $3$ under the first scenario for the random and directional profiles, respectively. We observe that the positioning performance is mainly affected by the distance between UE $3$ and the RIS in both figures. The results of the second scenario are shown in Figures~\ref{fig:heat_move_UE3}(c) and \ref{fig:heat_move_UE3}(d) for both profiles. In comparison to Figures~\ref{fig:heat_move_UE3}(a) and (b), Figures~\ref{fig:heat_move_UE3}(c) and (d) show slightly better performance in terms of the PEB which is due to the placement of UE $2$ further away from UE $1$ which results in better coverage. Figures~\ref{fig:heat_move_UE3}(e) and \ref{fig:heat_move_UE3}(f) shows the results of the third scenario. In comparison to Figures~\ref{fig:heat_move_UE3}(a) and (b), Figures~\ref{fig:heat_move_UE3}(e) and (f) show worse positioning performance which happens because UE $3$ is placed at a higher height (i.e., larger distance from the RIS and both UEs). As expected, the directional profile outperforms the random one in the three scenarios. RIS-aided localization systems might suffer from blind areas \cite{JrCUP, blind_area}. However, Figure~\ref{fig:heat_move_UE3} shows that the proposed use case does not suffer from blind areas and the localization error is mainly affected by the distance from the RIS and the other UEs.

\subsection{Effect of Multi-path on Positioning Performance}
\label{sec:SP simulation}
In this section, we study the effect of the multi-path on the positioning performance. We consider the same setup as in Section~\ref{sec:simulation_setup} and we generate four SPs for each receiving UE (i.e., $S_j=4~\forall j$) randomly in the environment. We also consider three different values for the RCS which are $\sigma^2_{\text{rcs}}=1,~10$ and $30$ $\text{m}^2$. Here, we consider $100$ realizations of the random codebook and we take the average of $200$ noise realizations for each codebook. For ease of illustration, we only consider the positioning error of UE $3$.

\begin{figure}[!t]
  \centering  
\centerline{\input{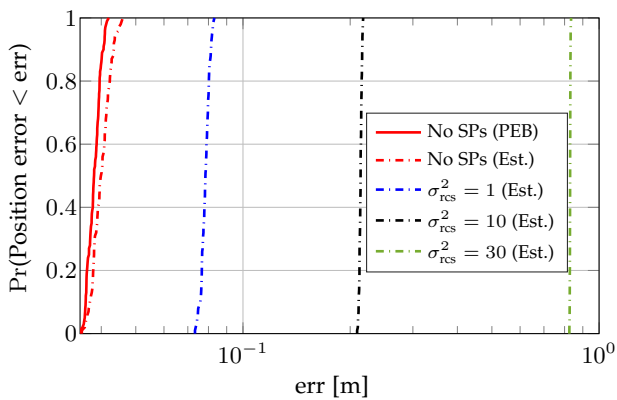}}
\caption{ECDF of the positioning error of UE $3$ under multi-path scenario with $S_j=4~\forall j$ and different RCS values.}
\label{Fig:effect_of_sp} \vspace{-5mm}
\end{figure}

In Figure~\ref{Fig:effect_of_sp}, the solid red curve represents the ECDF of the PEB while the dashed red curve represents the ECDF of the RMSE of the estimator with zero SPs involved. The positioning performance degrades if there are SPs in the environment and gets worse as the RCS increases. However, even with $\sigma^2_{\text{rcs}}=30$ $\text{m}^2$, sub-meter level positioning can still be achieved.    

\subsection{Effect of Number of UEs on Positioning Performance}
In this section, we study the effect of the number of UEs involved in the localization algorithm on  the positioning performance for two RIS profiles. Here, there are $121$ reflecting elements at the RIS and the number of transmissions per UE is $40$. The results are averaged over $20$ different codebook realizations. For the directional profile, we set $\Sigmam_{\pv_{\text{U}_i}}=1.5\mathbf{I}_3$

\begin{figure}[!t]
  \centering  
\centerline{
%
%
\begin{tikzpicture}

\begin{axis}[%
width=65mm,
height=42mm,
at={(0mm,0mm)},,
scale only axis,
xmin=3,
xmax=6,
xtick={3, 4, 5, 6},
xticklabel style = {font=\small,yshift=0.5ex},
yticklabel style = {font=\small,xshift=0ex},
ymode=log,
ymin=0.00711785229657392,
ymax=0.118990217406347,
yminorticks=true,
axis background/.style={fill=white},
xmajorgrids,
ymajorgrids,
legend style={font=\scriptsize, at={(1, 1)}, anchor=north east,legend cell align=left, align=left, draw=white!15!black}
]
\addplot [color=red, line width=1.0pt, mark=asterisk, mark options={solid, red}]
  table[row sep=crcr]{%
3	0.118990217406347\\
4	0.0650002154542546\\
5	0.0221312414126861\\
6	0.0198433598753049\\
};
\addlegendentry{random codebook}


\addplot [color=black, line width=1.0pt, mark=+, mark options={solid, black}]
  table[row sep=crcr]{%
3	0.0345527312381957\\
4	0.0202285474748619\\
5	0.00925873776614177\\
6	0.00811785229657392\\
};
\addlegendentry{directional codebook}

\end{axis}
\node[rotate=0,fill=white] (BOC6) at (3.35cm,-.7cm){\small Number of UEs};
\node[rotate=90] at (-8mm,22mm){\small PEB [m]};
\end{tikzpicture}%
}
\caption{The PEB for UE $1$ for different RIS profiles and different number of UEs involved in the localization.}
\label{Fig:number_of_ues} \vspace{-5mm}
\end{figure}
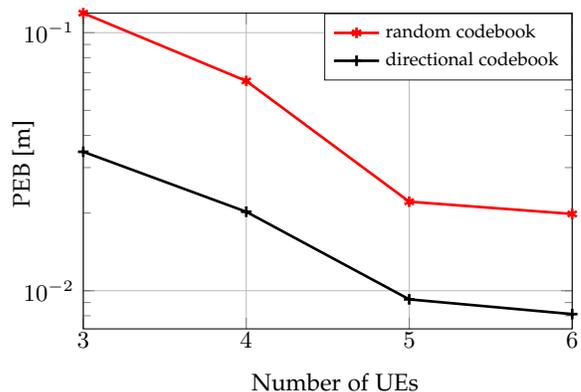

Figure~\ref{Fig:number_of_ues} shows the PEB of UE $1$ for random and directional RIS profiles as the number of involved UEs increases. Here, we generate the positions of the UEs randomly over the area $[1,11]$ m, $[-6, 6]$ m and $[-2,2]$ m on the $x$-axis, $y$-axis and $z$-axis, respectively. We can see that the best positioning performance is provided by the directional codebook, followed by the random phase codebook. The localization accuracy improves by involving more UEs, but that will increase the computational complexity since more measurements are needed. For example, the total number of transmissions is $120$ (around $1$ ms in duration) in the case of $3$ UEs while the total number of transmissions is 240 (around $2$ ms) for $6$ UEs.  
 
\section{Conclusion}
\label{sec:conclusion}
In this paper, we studied a 3D CP scenario with one RIS and $K$ UEs in order to estimate the locations of the UEs without using a BS via sidelink communications. We started by introducing the channel model and deriving the lower bounds of the channel parameters and position estimates. We then introduced two practical RIS profiles and formulated a power allocation optimization problem for fixed RIS profile to minimize the average PEB. We also developed three low-complexity estimators to estimate the time delays, spatial frequencies, and positions of the UEs. Through numerical studies, we showed that a sufficiently large RIS could replace an AP without any performance degradation. The proposed localization algorithm can achieve sub-meter level accuracy even under multi-path scenarios. We also evaluated the positioning performance under different RIS profiles and showed that directional profiles resulted in the best performance when the uncertainty of the prior information was low and the power allocation was optimal.

\allowdisplaybreaks
\appendices
\section{Derivatives of $\muv_{ij,t}$ }
\label{partial_signal_der}
The partial derivatives of $\muv_{ij,t}$ with respect to $\eta_b$, $b=1,2, \ldots, 8(K-1)K$, can be computed as:
\begin{align}
        &\frac{\partial \muv_{ij,t}}{\partial \tau_{ij}} = -j2\pi \Delta f  \sqrt{E_i} \beta_{ij} \left( \nv\odot \dv\left(\tau_{ij}\right) \right), \label{partial-derivatives-first}\\
        &\frac{\partial \muv_{ij,t}}{\partial \tau_{ij,r}} = -j2\pi \Delta f  \sqrt{E_i} \beta_{ij,r} \left( \nv\odot \dv\left(\tau_{ij,r}\right) \right)\cv^\top\left(\boldsymbol{\gamma}_{ij}\right)  \omegav_{i,t},\\
        &\frac{\partial \muv_{ij,t}}{\partial \xi_{ij}} = j\frac{2\pi}{\lambda} \sqrt{E_i} \beta_{ij,r} \dv\left(\tau_{ij,r}\right) \left(\pv_y \odot\cv\left(\boldsymbol{\gamma}_{ij}\right)\right)^\top  \omegav_{i,t},\\
        &\frac{\partial \muv_{ij,t}}{\partial \zeta_{ij}} = j\frac{2\pi}{\lambda} \sqrt{E_i} \beta_{ij,r} \dv\left(\tau_{ij,r}\right) \left(\pv_z \odot\cv\left(\boldsymbol{\gamma}_{ij}\right)\right)^\top  \omegav_{i,t},\\
        &\frac{\partial \muv_{ij,t}}{\partial \alpha_{ij}} = \sqrt{E_i} e^{j\rho_{ij}}\dv\left(\tau_{ij}\right) ,\\
        &\frac{\partial \muv_{ij,t}}{\partial \rho_{ij}} = j \sqrt{E_i} \beta_{ij} \dv\left(\tau_{ij}\right) ,\\
        &\frac{\partial \muv_{ij,t}}{\partial \alpha_{ij,r}} = \sqrt{E_i} e^{j\rho_{ij,r}} \dv\left(\tau_{ij,r}\right) \cv^\top\left(\boldsymbol{\gamma}_{ij}\right)  \omegav_{i,t}, \\
        &\frac{\partial \muv_{ij,t}}{\partial \rho_{ij,r}} = j \sqrt{E_i} \beta_{ij,r}\dv\left(\tau_{ij,r}\right) \cv^\top\left(\boldsymbol{\gamma}_{ij}\right)  \omegav_{i,t}, 
        \label{partial-derivatives-last} 
      \end{align}  
where $\nv= \begin{bmatrix} 0, 1, \!\!& \!\!  \ldots, \!\!\! & N-1 \end{bmatrix} ^T$. 
$\pv_y= \begin{bmatrix} p_{1,y}, p_{2,y}, \!\!& \!\!  \ldots, \!\!\! & p_{M,y} \end{bmatrix} ^T$ where $p_{m,y}$ represents the location of the $m$-th RIS element along the $y$-axis. Similarly, $\pv_z= \begin{bmatrix} p_{1,z}, p_{2,z}, \!\!& \!\!  \ldots, \!\!\! & p_{M,z} \end{bmatrix} ^T$ denotes the RIS elements positions' along the $z$-axis. Please note that the partial derivatives in \eqref{partial-derivatives-first}--\eqref{partial-derivatives-last} will be equal to the zero vector if the indices on the numerator and denominator are different. For example, $\frac{\partial \muv_{ij,t}}{\partial \tau_{bv}}=\boldsymbol{0}_N$ if either $i\neq b$ or $j\neq v$ where $\boldsymbol{0}_N$ is a zero column vector of length $N$.

\section{Derivatives of $\etav$ }
\label{partial_eta}
The partial derivatives of the channel parameters with respect to $\pv_{\text{U}_k}$ (i.e., UE $k$) can be computed as:
\begin{align}
        \begin{split} 
        \frac{\partial \tau_{ij}}{\partial \pv_{\text{U}_k} } &= \frac{\partial \tau_{ji}}{\partial \pv_{\text{U}_k} }=  \\ &\begin{cases}
        \frac{1}{cd_{\text{U}_{k}\text{U}_{j}}} \left(\pv_{\text{U}_k}-\pv_{\text{U}_j}\right)& \text{if}~k=i \\
        \frac{1}{cd_{\text{U}_{i}\text{U}_{k}}} \left(\pv_{\text{U}_k}-\pv_{\text{U}_i}\right)& \text{if}~k=j \\
         \begin{bmatrix}  0, 0,0  \end{bmatrix}^\top& \text{otherwise},
        \end{cases}
        \end{split}\\
        \begin{split}
        \frac{\partial \tau_{ij,r}}{\partial \pv_{\text{U}_k} } &= \frac{\partial \tau_{ji,r}}{\partial \pv_{\text{U}_k} }=  \\ &\begin{cases}
        \frac{1}{cd_{\text{R}\text{U}_{k}}} \pv_{\text{U}_k}& \text{if}~k=i~\text{or}~k=j \\
         \begin{bmatrix}  0, 0,0  \end{bmatrix}^\top& \text{otherwise}.
        \end{cases}
        \end{split}
\end{align}
The derivatives of the spatial frequencies with respect to $\pv_{\text{U}_k}$ can be computed as:
\begin{align}
        &\frac{\partial \xi_{ij}}{\partial \pv_{\text{U}_k}} = \nonumber\\ &~~\begin{cases}
        \begin{bmatrix}  -\frac{x_{\text{U}_k} y_{\text{U}_k}}{d_{\text{R}\text{U}_k}^3}, \frac{1}{d_{\text{R}\text{U}_k}}-\frac{y_{\text{U}_k}^2}{d_{\text{R}\text{U}_k}^3},-\frac{z_{\text{U}_k} y_{\text{U}_k}}{d_{\text{R}\text{U}_k}^3}  \end{bmatrix}^\top &\text{if}~k=i~\text{or}~k=j\\
        \begin{bmatrix}  0, 0,0  \end{bmatrix}^\top& \text{otherwise},
        \end{cases}
\end{align}
\begin{align}
        &\frac{\partial \zeta_{ij}}{\partial \pv_{\text{U}_k}} = \nonumber\\ &~~\begin{cases}
        \begin{bmatrix}  -\frac{x_{\text{U}_k} z_{\text{U}_k}}{d_{\text{R}\text{U}_k}^3}, -\frac{y_{\text{U}_k} z_{\text{U}_k}}{d_{\text{R}\text{U}_k}^3}, \frac{1}{d_{\text{R}\text{U}_k}}-\frac{z_{\text{U}_k}^2}{d_{\text{R}\text{U}_k}^3} \end{bmatrix}^\top &\text{if}~k=i~\text{or}~k=j\\
        \begin{bmatrix}  0, 0,0  \end{bmatrix}^\top& \text{otherwise}.
        \end{cases}
\end{align}
The derivatives of the channel gains with respect to $\pv_{\text{U}_k}$ is the zero vector. In other words, 
\begin{equation}
\begin{split}
    \frac{\partial \alpha_{ij}}{\partial \pv_{\text{U}_k}}= \frac{\partial \alpha_{ij,r}}{\partial \pv_{\text{U}_k}} =  \frac{\partial \rho_{ij}}{\partial \pv_{\text{U}_k}} = \frac{\partial \rho_{ij,r}}{\partial \pv_{\text{U}_k}} =
    \begin{bmatrix}
        0,0,0
    \end{bmatrix}^\top \forall~i,~j,~k.
    \end{split}
\end{equation}
Next, the derivatives of the channel parameters with respect to the clock offset $\Delta_{t_k}$ can be computed as:
\begin{align}
\begin{split}
    \frac{\partial \tau_{ij}}{\partial \Delta_{t_k}} &= \frac{\partial \tau_{ij,r}}{\partial \Delta_{t_k}} = \begin{cases}
    1 &\text{if}~k=j\\
    -1 & \text{if}~k=i\\
    0 &\text{otherwise},
    \end{cases}
\end{split}\\
\begin{split}
    \frac{\partial \xi_{ij}}{\partial \Delta_{t_k}} &= \frac{\partial \zeta_{ij}}{\partial \Delta_{t_k}} = \frac{\partial \alpha_{ij}}{\partial \Delta_{t_k}}= \frac{\partial \alpha_{ij,r}}{\partial \Delta_{t_k}}  \\ &=\frac{\partial \rho_{ij}}{\partial \Delta_{t_k}} = \frac{\partial \rho_{ij,r}}{\partial \Delta_{t_k}} =
    0 ~ \forall~i,~j,~k.
    \end{split}
\end{align}
\ifCLASSOPTIONcaptionsoff
  \newpage
\fi



%
\bibliographystyle{IEEEtran}
\bibliography{main}
\vskip 0pt plus -1fil
\begin{IEEEbiography}
[{\includegraphics[width=1in,height=1.25in,clip,keepaspectratio]{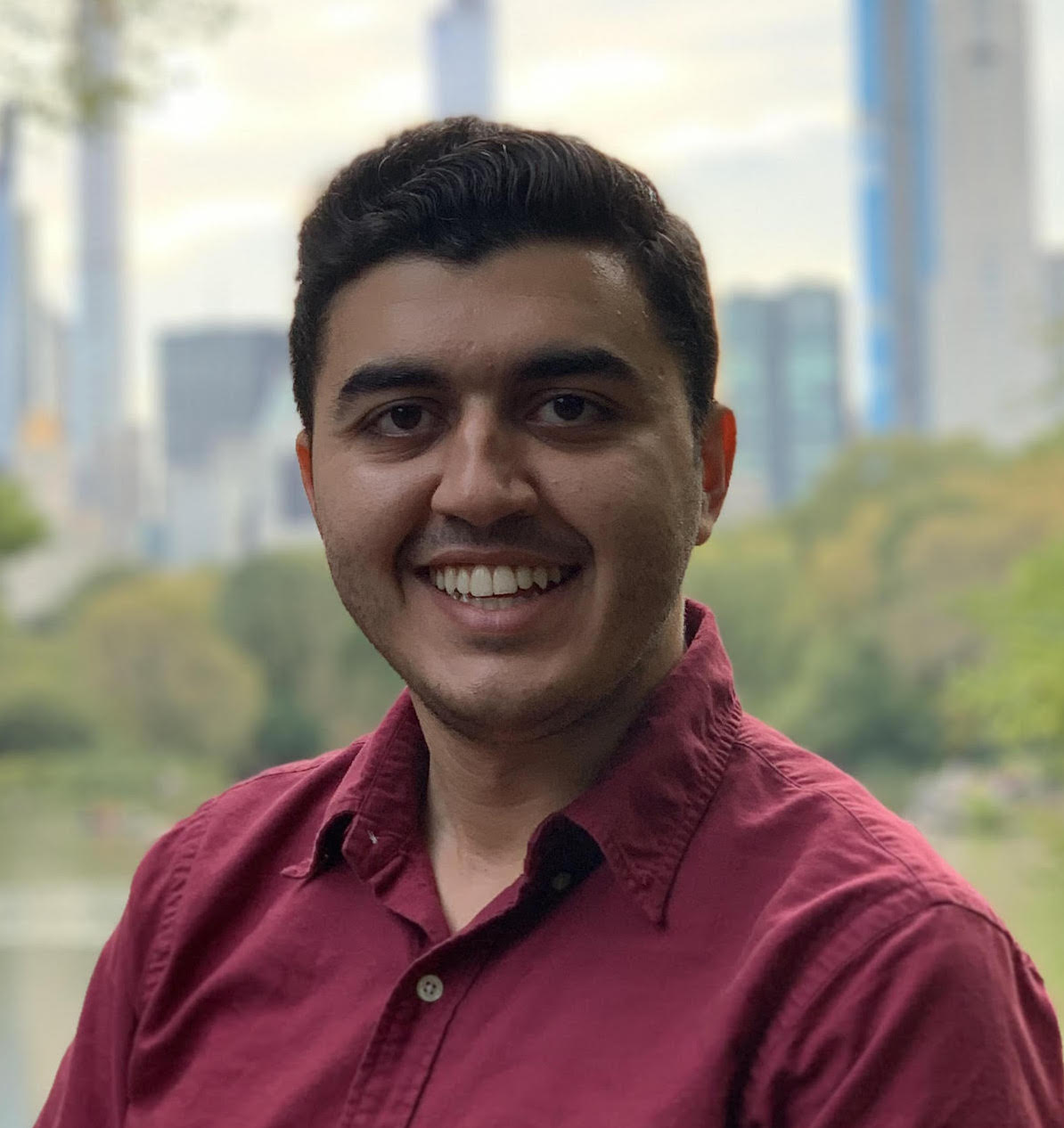}}]{Mustafa~Ammous} (S'17) received his B.Sc. degree in Electrical Engineering from King Fahd University of Petroleum and Minerals, Dhahran, Saudi Arabia in 2016 and his M.Sc. degree in Electrical and Computer Engineering at University of Idaho, Moscow, Idaho, United States in 2018. He is currently pursuing his PhD degree at University of Toronto, Toronto, Ontario, Canada. His research interests include radio localization and sensing using reconfigurable intelligent surfaces.
\end{IEEEbiography}
\vskip 0pt plus -1fil
\begin{IEEEbiography}
[{\includegraphics[width=1in,height=1.25in,clip,keepaspectratio]{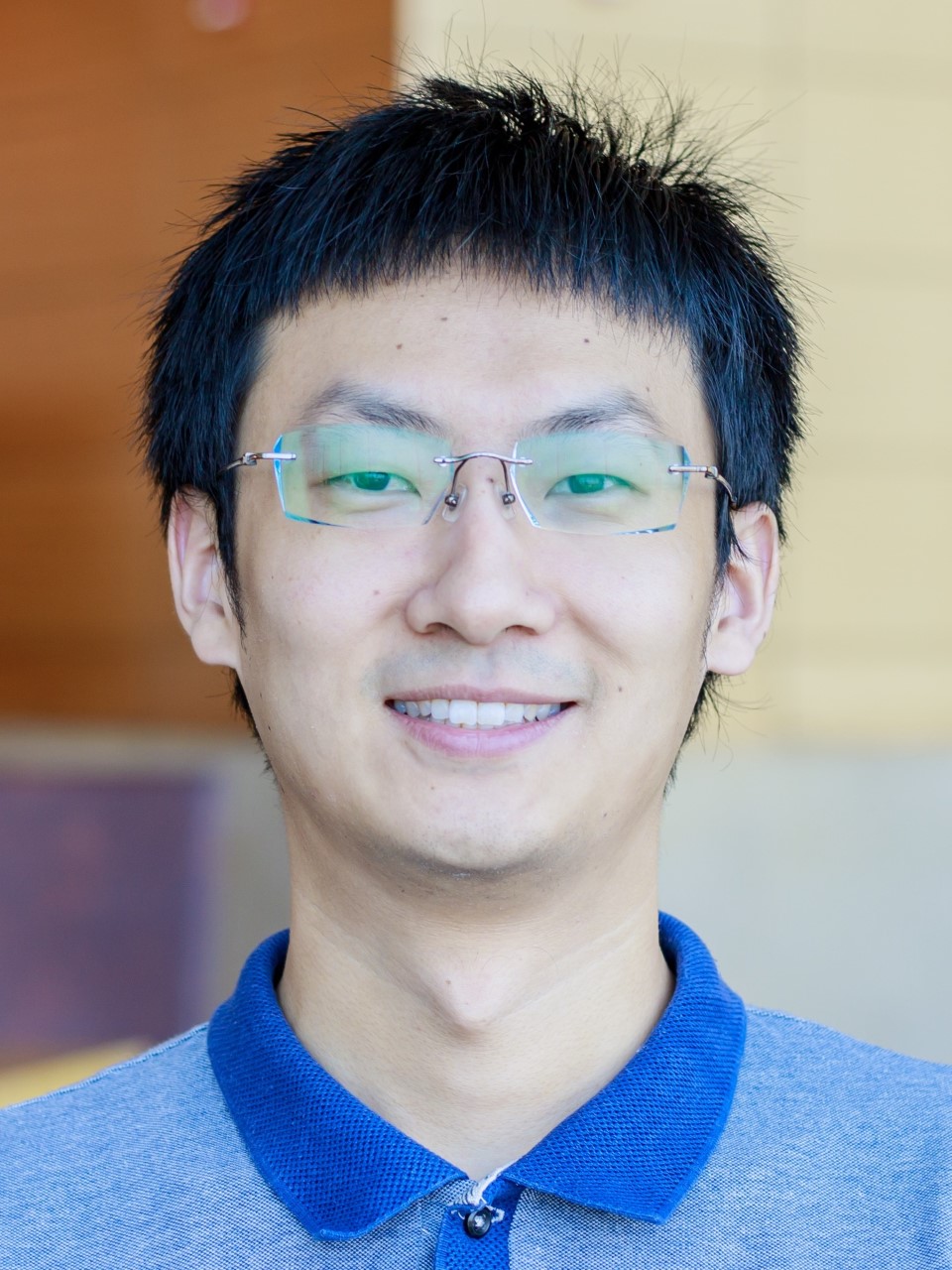}}]{Hui~Chen} (S'18, M'21) obtained the Ph.D. degree in electrical and computer engineering from the King Abdullah University of Science and Technology (KAUST), Thuwal, Saudi Arabia, in 2021. He is currently a Post-Doctoral Researcher with the Chalmers University of Technology, Gothenburg, Sweden. His research interests include 5G/6G (mmWave/THz) localization and sensing, stochastic optimization, and machine learning for signal processing.
\end{IEEEbiography}
\vskip 0pt plus -1fil

\begin{IEEEbiography}[{\includegraphics[width=1in,height=1.25in,clip,keepaspectratio]{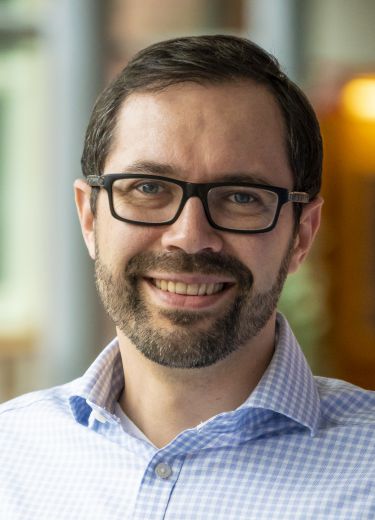}}]{Henk~Wymeersch}
	(S'01, M'05, SM'19) obtained the Ph.D. degree in Electrical Engineering/Applied Sciences in 2005 from Ghent University, Belgium. He is currently a Professor of Communication Systems with the Department of Electrical Engineering at Chalmers University of Technology, Sweden. He is also a Distinguished Research Associate with Eindhoven University of Technology. Prior to joining Chalmers, he was a postdoctoral researcher from 2005 until 2009 with the Laboratory for Information and Decision Systems at the Massachusetts Institute of Technology. Prof. Wymeersch served as Associate Editor for IEEE Communication Letters (2009-2013), IEEE Transactions on Wireless Communications (since 2013), and IEEE Transactions on Communications (2016-2018) and is currently Senior Member of the IEEE Signal Processing Magazine Editorial Board.  During 2019-2021, he was an IEEE Distinguished Lecturer with the Vehicular Technology Society.  His current research interests include the convergence of communication and sensing, in a 5G and Beyond 5G context. 
\end{IEEEbiography}
\vskip 0pt plus -1fil

\begin{IEEEbiography}
[{\includegraphics[width=1in,height=1.25in,clip,keepaspectratio]{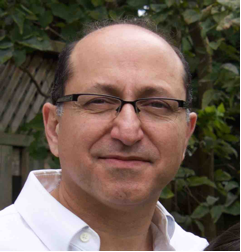}}]{Shahrokh ~Valaee } is a Professor with the Edward S. Rogers Sr. Department of Electrical and Computer Engineering, University of Toronto, and the holder of Nortel Chair of Network Architectures and Services. He is the Founder and the Director of the Wireless Innovation Research Laboratory (WIRLab) at the University of Toronto. Professor Valaee was the TPC Co-Chair and the Local Organization Chair of the IEEE Personal Mobile Indoor Radio Communication (PIMRC) Symposium 2011. He was the TPC Co-Chair of ICT 2015, and PIMRC 2017, and the Track Co-Chair of WCNC 2014, PIMRC 2020, VTC Fall 2020. He is the co-chair of the organizing committee for PIMRC 2023. From December 2010 to December 2012, he was the Associate Editor of the IEEE Signal Processing Letters. From 2010 to 2015, he served as an Editor of IEEE Transactions on Wireless Communications. Currently, he is an Editor of the Journal of Computer and System Science and serves as a Distinguished Lecturer for IEEE Communication Society. He was the co-recipient of the best paper award in the IEEE Machine Learning for Signal Processing (MLSP) 2020 workshop. Professor Valaee is a Fellow of the Engineering Institute of Canada, and a Fellow of IEEE. 
\end{IEEEbiography}

\end{document}